\documentclass{article}
\PassOptionsToPackage{round}{natbib}
\usepackage[main,final]{neurips_2026}

\usepackage[utf8]{inputenc}
\usepackage[T1]{fontenc}
\usepackage{hyperref}
\usepackage{url}
\usepackage{booktabs}
\usepackage{amsfonts}
\usepackage{nicefrac}
\usepackage{microtype}
\usepackage{xcolor}

\usepackage{amssymb,amsthm,amsmath,mathtools}
\usepackage{thmtools,thm-restate}
\usepackage{pifont}

\newif\ifverbose

\usepackage{algorithm,algpseudocode}

\usepackage{graphicx}
\usepackage{subcaption}
\usepackage[labelfont={normalfont,bf},textfont=it]{caption}
\usepackage{tabularx}
\usepackage{array,multirow,bigdelim}

\usepackage{pgf,tikz}
\usepackage{tikz-cd}
\usetikzlibrary{fit,calc,math,shapes,decorations.text,arrows,
  decorations.markings,decorations.pathmorphing,shapes.geometric,
  positioning,decorations.pathreplacing}
\tikzset{snake it/.style={decorate, decoration=snake}}

\usepackage{enumerate}
\usepackage{etoolbox}
\usepackage{dsfont}
\usepackage{mathrsfs,amsfonts}
\usepackage[capitalise,noabbrev,nameinlink]{cleveref}
\Crefname{property}{Property}{Properties}
\Crefname{theorem}{Theorem}{Theorems}
\Crefname{example}{Example}{Examples}
\Crefname{table}{Table}{Tables}
\Crefname{algorithm}{Algorithm}{Algorithms}
\Crefname{figure}{Figure}{Figures}
\Crefname{lemma}{Lemma}{Lemmas}
\Crefname{definition}{Definition}{Definitions}
\usepackage{cancel}
\usepackage{enumitem}
\definecolor{DarkGreen}{rgb}{0.1,0.5,0.1}
\colorlet{myorange}{orange!40}

\DeclareMathOperator*{\argmax}{\arg\!\max}

\newcommand{\basis}{\textsc{Basis}}
\newcommand{\impr}{\textsc{Improvize}}
\newcommand{\nl}{\textsc{null}}
\newcommand{\clAlg}{\textsc{SeqShare}}
\newcommand{\ExtSh}{\textsc{Ext-Shap}}
\newcommand{\MargSol}{\textsc{MargSol}}
\newcommand{\SeqSh}{\textsc{SeqSh}}
\newcommand{\seqgame}{$\langle N,v \rangle$}
\newcommand{\swap}[2]{\textup{swap}(#1,#2)}

\newcommand\ddfrac[2]{\frac{#1}{#2}}

\newtheorem{theorem}{Theorem}
\newtheorem{lemma}{Lemma}

\newtheorem{prop}{Proposition}
\theoremstyle{definition}

\newtheorem{examplex}{Example}

\newenvironment{example}{\pushQED{\qed}\examplex}{\popQED\endexamplex}
\theoremstyle{remark}

\newtheorem{definition}{Definition}

\newenvironment{psketch}{%
  \par\noindent\textit{Proof Sketch.}\hspace*{0.5em}%
}{\hfill$\square$\par}

\allowdisplaybreaks

\title{Sequential Solution Concepts in Cooperative Games\\
with Generalized Characteristic Functions}

\author{%
  {Ashwin Goyal} \\
  {IIT Bombay} \\
  \texttt{ashwin03goyal@gmail.com} \\
  \And
  {Drashthi Doshi} \\
  {IIT Bombay} \\
  \texttt{drashthi@cse.iitb.ac.in} \\
  \And
  {Swaprava Nath} \\
  {IIT Bombay} \\
  \texttt{swaprava@cse.iitb.ac.in} \\
}

\begin{document}

\maketitle

\begin{abstract}
Motivated by the fact that the worth of a coalition may depend on the order in which agents arrive, \citet{NOWAK1994150} introduced cooperative games
with generalized characteristic functions, where the worth function $v$ is
defined on sequences of agents $\pi$ rather than sets $S$.  We call such
games \emph{temporal cooperative games (TCGs)}.  This order sensitivity
necessitates a re-examination of axioms for reward sharing.  Prior work
proposed several axioms; however, the resulting solution concepts are
inherently order-oblivious and closely tied to the Shapley value.  In
contrast, we focus on sequential solution concepts that explicitly depend on
the realized order $\pi$.  We study reward-sharing mechanisms satisfying
\emph{incentive for optimal arrival} (I4OA), which promotes orders that
maximize total worth; \emph{online individual rationality} (OIR), which
ensures later arrivals do not harm existing agents; and \emph{sequential
efficiency} (SE), which requires that the worth of any sequence is fully
distributed among its agents.  These axioms are intrinsic to TCGs, and we
characterize a class of reward-sharing mechanisms uniquely determined by
them.  The classical Shapley value does not directly extend to this setting.
We therefore construct natural Shapley analogs in two worlds: a
\emph{sequential} world, where rewards are defined for each sequence--agent
pair, and an \emph{extended} world, where rewards are defined per agent,
consistent with the framework of \citet{NOWAK1994150}.  In both cases, the
axioms of efficiency, additivity, and null player uniquely characterize the
corresponding Shapley analogs.  However, these Shapley analogs are disjoint
from the class of solutions satisfying the sequential axioms, even for
convex and simple TCGs.  Our results reveal a fundamental tension in
temporal cooperative games: when order matters, solution concepts satisfying
natural sequential incentives must differ structurally from Shapley-based
allocations, motivating new axiomatic foundations for sequential reward
sharing.
\end{abstract}

\section{Introduction}

Consider hospitals collaborating to train a medical image classifier for a rare progressive disease via \emph{federated learning}.
Hospital~A holds early-stage images with subtle texture changes;
Hospital~B holds mid-stage images with visible structural deformations;
Hospital~C holds late-stage images where the disease is overt.
Each hospital's gradient update shapes the loss landscape that subsequent participants optimize over.
If Hospital~C joins first, the model anchors on coarse structural signals,
and Hospital~A's subtle texture updates later struggle to move it.
The result is a classifier that misses early-stage diagnoses, which are clinically most valuable.
Reverse the order and the model builds rich low-level representations first, with each subsequent participant adding complementary structure.
The same three participants, the same data, the same algorithm, but a different joining order yields a substantially better model.
The same phenomenon arises in software development pipelines, robotic assembly, and supply chains: the order in which agents join determines the value created, not just who joins.

Classical cooperative games~\citep{maschler2020game} overlook this: their
worth%
\footnote{We use the terms \emph{worth} and \emph{characteristic} functions interchangeably, as is common in the literature, and denote
them by $v$.}
functions are defined only for coalitions, ignoring the order of agents.
Early works by \citet{NOWAK1994150,sanchez1997values} incorporate order
through \emph{generalized characteristic functions} (GCF), and follow-up
papers~\citep{van2014ordermonotonic,feng2013matrix,yuan2013consistency,
Li2024restricted,michalak2014implementation,zou2020extended} extend their
axioms---but all yield \emph{order-oblivious} solution concepts where rewards do not depend on the arrival order.
We focus on \emph{order-sensitive} sequential solution concepts and study three
natural axioms for them.
\emph{Online individual rationality} (OIR) ensures Hospital~A is not penalized after later arrivals improve the model, preserving its
incentive to join early.
\emph{Incentive for optimal arrival} (I4OA) incentivizes each hospital to join in the order that maximizes diagnostic accuracy.
\emph{Sequential efficiency} (SE) ensures the full value of any realized sequence is distributed among its participants.
We call cooperative games with generalized characteristic functions
\emph{temporal cooperative games} (TCGs).

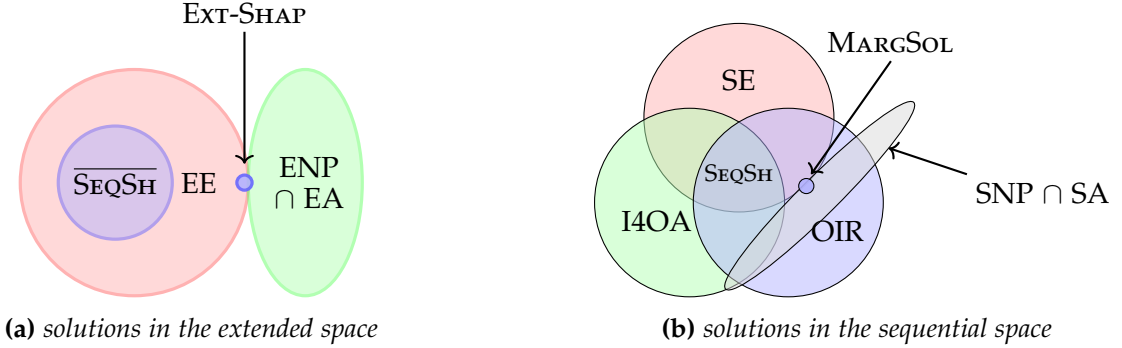
\begin{figure}[t]
    \centering
    \begin{subfigure}[t]{0.45\linewidth}
        \centering
        \scalebox{0.99}{
        \begin{tikzpicture}
            \filldraw[color=red!60, fill=red!30, very thick,opacity=0.5](-1,0) circle (1.5);
            \filldraw[color=blue!60, fill=blue!30, very thick,opacity=0.5](-1.25,0) circle (0.75);
            \filldraw[color=green!60, fill=green!30, very thick,opacity=0.5](1.25,0) ellipse (0.75 and 1.5);
            \filldraw[color=blue!60, fill=blue!30, very thick,opacity=0.9](0.45,0) circle (0.1);
            \draw[->, thick] (0.45,2) -- (0.45,0.25);
            \node at (0.45,2.25) {\ExtSh{}};
            \node at (1.3,0.2) {ENP};
            \node at (1.3,-0.2) {$\cap$ EA};
            \node at (-1.25,0) {$\overline{\SeqSh}$};
            \node at (-0.15,0) {EE};
        \end{tikzpicture}}
        \caption{Solutions in the extended space.}
        \label{fig:extended}
    \end{subfigure}
    \hspace{0.005\linewidth}
    \begin{subfigure}[t]{0.45\linewidth}
        \centering
        \scalebox{0.99}{
        \begin{tikzpicture}
            \draw[fill=red!30, fill opacity=0.5] (90:0.75) circle (1.25);
            \draw[fill=green!30, fill opacity=0.5] (210:0.75) circle (1.25);
            \draw[fill=blue!30, fill opacity=0.5]  (330:0.75) circle (1.25);
            \draw[fill=gray!30, fill opacity=0.5, rotate=45]  (300:1.1) ellipse (1.75 and 0.25);
            \draw[->, thick] (3,0) -- (10:2);
            \node at (3.25,-0.25) {SNP $\cap$ SA};
            \draw[fill=blue!30, fill opacity=0.9]  (350:0.9) circle (0.1);
            \draw[->, thick] (2,1.5) -- (360:0.95);
            \node at (2,1.75){\MargSol{}};
            \node at (90:1.25) {SE};
            \node at (210:1.25) {I4OA};
            \node at (330:1.5) {OIR};
            \node at (330:0) {\footnotesize{\SeqSh}};
        \end{tikzpicture}}
        \caption{Solutions in the sequential space.}
        \label{fig:sequential}
    \end{subfigure}
    \caption{Impossibility of Shapley-inspired properties and solutions in
      \clAlg{} (\SeqSh{}).}
    \label{fig:combined}
\end{figure}

Ideally, $\phi$ would also satisfy Shapley fairness by giving each hospital its average marginal contribution across all joining orders, which is the standard notion of equitable reward sharing.
However, we prove this is \emph{impossible}: sequential solution concepts satisfying
OIR, I4OA, and SE are provably disjoint from those satisfying
Shapley-inspired properties such as efficiency, additivity, and the
null-player condition~(\Cref{fig:combined}), even for convex and simple
TCGs.
No reward rule can simultaneously protect early participants, incentivize the optimal joining order, and distribute Shapley-fair shares---a fundamental constraint for any mechanism designer in sequential collaborative settings.

\subsection{Our Contributions}

\begin{itemize}[leftmargin=14pt, itemsep=0pt, parsep=0pt, topsep=0pt]
  \item We introduce \emph{sequential} solution concepts
    (\Cref{sec:prelim}).  Unlike existing GCF literature, which yields
    order-oblivious reward shares, our framework assigns rewards to every
    (sequence, player) pair, arguing that order-sensitive environments require
    \emph{order-sensitive} solution concepts.

  \item We formalize axioms natural to this order-sensitive setting: I4OA,
    OIR, and SE (\Cref{sec:prelim}).

  \item We characterize all sequential solution concepts satisfying I4OA, OIR, and SE
    via a class of reward-sharing mechanisms \clAlg{} (\Cref{algo:SeqSh}),
    and identify a necessary condition, termed \emph{basis}, for their
    satisfaction (\Cref{lem:I4OA_efficiency_and_basis}).

  \item We formulate order-sensitive analogs of Shapley's axioms for TCGs:
    efficiency, additivity, and null player (\Cref{sec:shapley-prop}), and
    show they uniquely characterize a Shapley-analogous sequential solution
    concept \MargSol{} (\Cref{sec:unique-margsol}); symmetry follows as a
    consequence (\Cref{sec:symmetry-not-necessary}).

  \item To connect with order-oblivious solutions, we define \emph{extended}
    solution concepts by averaging over sequences.  The extended axioms
    uniquely identify \ExtSh{} (\Cref{sec:unique-extsh}), aligning exactly
    with existing GCF literature.

  \item Finally, we show that the extended solutions induced by \clAlg{} are
    disjoint from \ExtSh{}, establishing a fundamental incompatibility between
    desirable sequential properties and Shapley-inspired axioms, even for
    \emph{convex} and \emph{simple} games (\Cref{sec:bestofboth}).
\end{itemize}

\subsection{Related Work}
\label{sec:literature}

Classical cooperative game theory originated with \citet{von1947theory}, with foundational solution concepts including the core \citep{gillies1959solutions}, nucleolus \citep{schmeidler1969nucleolus}, and Shapley value \citep{Shapley1953}. Characterizations of the Shapley value via variants of its axioms have been studied extensively \citep{van2002axiomatizationshapley, casajus2013nullshapley, hamiache2001associated, casajus2014shapley}. Dynamic and online extensions include coalition formation under random arrivals or dissolution \citep{bullinger2023onlineonlinecoalitionrandom, bullinger2025stabilityonlineformation}, general online coalition formation \citep{flammini2021onlinecoalition}, and cooperative games where worth evolves over time \citep{lehrer2013core, habis2010note, kranich2005core}. Online cooperative games with sequential arrivals have been studied for monotone games \citep{zhangearlyarrivalcooperative}, cost-sharing \citep{zhaocostsharing}, axiomatic frameworks \citep{aziz2025participation}, and stable coalition formation \citep{zhang2025coalitionsflycooperativegames}; however, all these models define worth over coalitions rather than sequences.

Generalized characteristic functions were introduced by \citet{NOWAK1994150} and extended by \citet{sanchez1997values, sanchez1999coalitionalgeneralized, bergantinos2001weightedGeneralized}. Despite allowing order-dependent worth, the resulting NR and SB solution concepts are sequence-independent. \citet{NOWAK1994150} extended Shapley-style axioms to TCGs and uniquely characterized a solution, while \citet{sanchez1997values} weakened the null-player property but required an additional symmetry assumption to recover uniqueness (see \Cref{sec:shapley-prop}), which we do not impose.

Subsequent work characterizes the SB value via bargaining \citep{yuan2013consistency} and consistency properties \citep{feng2013matrix}.   \citet{van2014ordermonotonic} characterize a class of solutions including the SB and NR values as extreme cases.  \citet{zou2020extended} and
\citet{Li2024restricted} study settings with restrictions on agent arrivals.  Finally, \citet{michalak2014implementation} implement
generalized games as perfect-information games that admit the NR and SB
values as subgame-perfect Nash equilibria.

The fundamental distinction between these works and ours lies in our focus on
order-sensitive sequential solution concepts.  This perspective leads to two key
departures: (1)~the axioms of interest differ and are motivated by settings
in which order-sensitive solutions are natural, and (2)~the required
properties are more fine-grained in the Shapley framework, as they must hold for
every sequence $\pi$.  Consequently, our main results are: (1)~a
characterization of order-sensitive axioms tailored to TCGs
(\Cref{thm:properties_iff_seqsh}), (2)~a characterization of Shapley-like
yet order-sensitive axioms (\Cref{Thm:MargSol_unique}), and (3)~a
demonstration of the disjointness of these two classes
(\Cref{sec:bestofboth}).

\section{Preliminaries}
\label{sec:prelim}

Consider a set of agents $N = \{1,\ldots,n\}$.  Let $\textsc{perm}(S)$
denote all permutations of players in $S \subseteq N$.  Define all possible
sequences of any length by $\Pi := \{\pi: \pi \in \textsc{perm}(S), S \in
2^N \setminus \emptyset\}$.  We denote the set of agents in a sequence $\pi$
by $P(\pi)$ and the player at position $i$ in $\pi$ as $\pi(i)$.  We define
a characteristic function $v: \Pi \to \mathbb{R}$, which assigns a worth to
every sequence of agents.  A \emph{temporal cooperative game (TCG)} is thus
described by a tuple \seqgame{}.  We also define $\Pi_{-N'} := \{\pi: \pi
\in \textsc{perm}(S), S \in 2^{N \setminus N'}\}$, i.e., the set of all
agent sequences excluding agents in $N'$.  Denote the last player of $\pi$
as $\ell(\pi)$.

We call $\pi'$ a \emph{prefix} of $\pi$ if the first $|P(\pi')|$ players of
$\pi$ coincide with those of $\pi'$, i.e., $\pi(i) = \pi'(i)$ for all $i =
1, 2, \ldots, |P(\pi')|$.  A prefix $\pi'$ of $\pi$ is denoted $\pi'
\sqsubset \pi$.  Note that $|P(\pi')| < |P(\pi)|$, otherwise they are the
same sequence.  We assume the characteristic function is \emph{monotone},
i.e., $v(\pi') \leqslant v(\pi)$ for all $\pi' \sqsubset \pi$.  We denote
the \emph{predecessor} of $i$ in $\pi$ by $\pi_i$, which is the longest
prefix of $\pi$ not containing $i$.  A sequence $\pi$ is a \emph{full
sequence} if $P(\pi)=N$.  Denote the set of all monotone characteristic
functions by $V$.
  The empty sequence $\emptyset$ is not an element of $\Pi$;
whenever an expression such as $v(\pi_i)$ refers to it (for instance, when
$i$ is the first agent of $\pi$), we use the convention $v(\emptyset)=0$.
We also assume that worths are nonnegative, $v(\pi) \geqslant 0$ for all
$\pi \in \Pi$, as in all of our motivating examples.

The term $\phi_i(\pi, v)$ denotes the reward of agent $i \in P(\pi)$ when
arriving in sequence $\pi$ in the TCG \seqgame{}.  We denote by
$\phi(\pi,v)$ the vector $(\phi_i(\pi,v), i \in N)$, with the convention
that $\phi_j(\pi,v) = 0$ for all $j \in N \setminus P(\pi)$.  We write
$\phi(\pi)$, dropping $v$ when clear from context.
A reward-sharing method is
\emph{nonnegative} if $\phi_i(\pi,v) \geqslant 0$ for all $\pi \in \Pi$ and
$i \in P(\pi)$; that is, no agent is ever asked to pay to participate.  For
example, in the federated learning setting of the introduction, a hospital contributes data and is never charged for doing so. The optimal sequence
$\pi^*(v) \in \argmax_{\pi \in \Pi} v(\pi)$ is defined with a tie-breaking
rule such that $P(\pi^*(v)) = N$ (with arbitrary tie-breaking among
full-length sequences), since the game is monotone.  Our results hold even
when $P(\pi^*(v)) \subset N$, and the proofs adapt easily.  We summarize the
notation in \Cref{tab:notation}.

\begin{table}[t]
\centering
\caption{Summary of Notation}
\label{tab:notation}
\small
\begin{tabularx}{\textwidth}{c>{\hsize=0.7\hsize}X@{\hspace{1em}}c>{\hsize=1.3\hsize}X}
\toprule
\textbf{Symbol} & \textbf{Description} & \textbf{Symbol} & \textbf{Description} \\
\midrule
$N = \{1, \ldots, n\}$ & Set of agents &
  $\pi_i$ & Predecessor of $i$: longest prefix of $\pi$ not containing $i$ \\
$2^N$ & Power set of $N$ &
  $v : \Pi \to \mathbb{R}$ & Characteristic function \\
$\textsc{perm}(S)$ & All permutations of $S \subseteq N$ &
  $V$ & Set of all monotone characteristic functions \\
$\Pi$ & All sequences over subsets of $N$ &
  $\langle N, v \rangle$ & A temporal cooperative game (TCG) \\
$\pi$ & A sequence of agents &
  $\Pi_{-N'}$ & All sequences over $N \setminus N'$ \\
$P(\pi)$ & Agents appearing in $\pi$ &
  $\pi^*(v)$ & Optimal sequence maximizing $v(\pi)$ \\
$\pi(i)$ & Agent at position $i$ in $\pi$ &
  $\phi_i(\pi, v)$ & Reward of agent $i$ in sequence $\pi$ under $v$ \\
$\ell(\pi)$ & Last player of $\pi$ &
  $\phi(\pi, v)$ & Reward vector for all agents in $\pi$ \\
$\pi' \sqsubset \pi$ & Proper prefix of $\pi$ & & \\
\bottomrule
\end{tabularx}
\end{table}

Since our focus is on reward-sharing mechanisms that incentivize agents to
arrive in the order that maximizes total worth, we first ask: under what
conditions can such incentives even be guaranteed?  Intuitively, for a
mechanism to incentivize every agent to join in the optimal sequence, there
must exist a consistent set of constraints on agent rewards and the worth of
every sequence.  We formalize this requirement as follows.

\begin{definition}[Basis Conditions]
\label{def:basis}
A TCG $\langle N, v \rangle$ satisfies the \emph{basis conditions} if there
exists a vector $x \in \mathbb{R}^n$ such that
\begin{equation}
\label{eq:basis}
  \textstyle
  \text{(i)}\ \sum_{i \in P(\pi)} x_i \geqslant v(\pi),\ \forall \pi \in \Pi,
  \qquad
  \text{(ii)}\ \sum_{i \in P(\pi^*(v))} x_i = v(\pi^*(v)).
\end{equation}
Any such $x$ is called a \emph{basis solution}.  A TCG admitting a basis
solution is called a \emph{basis-satisfiable TCG}. The basis conditions are analogous to the core of classical cooperative games.
\end{definition}

Before defining the axiomatic properties, we make the modelling assumptions
explicit.  We consider a complete-information setting: all agents in $N$ are
available, and the characteristic function $v$ is known.  The sequential solution concept $\phi$
is announced to all agents prior to any commitments, after which agents
choose when to arrive.  This is motivated by the \emph{incentive for early
arrival} (I4EA) property introduced by \citet{zhangearlyarrivalcooperative}:
agents are all available (e.g., candidates in a job market) but choose when
to commit, and a well-designed $\phi$ should make the optimal joining order
weakly dominant for every agent.

The players arrive sequentially in a TCG, so rewards must also be given as
they arrive.  When a new player $i$ arrives, extending sequence $\pi$ to
$\pi+i$, the worth generated must be distributed among the players present.

\begin{definition}[Online Individual Rationality]
\label{def:OIR}
A sequential solution concept $\phi$ satisfies \emph{online individual rationality}
(OIR) for a TCG $\langle N, v \rangle$ if $\phi_i(\pi,v) \leqslant
\phi_i(\pi',v)$ for all $i \in P(\pi)$ and all $\pi \sqsubset \pi'$.
\end{definition}

The sequential nature of the game also creates the possibility that players
stop arriving; the next player's arrival is uncertain.  Therefore, the
sequential solution concept should ensure the worth of any sequence $\pi$ is fully
distributed among its players.

\begin{definition}[Sequential Efficiency (SE)]
\label{def:SE}
A sequential solution concept $\phi$ satisfies \emph{sequential efficiency} (SE) for a
TCG $\langle N, v \rangle$ if $\sum_{i \in P(\pi)}\phi_i(\pi,v) = v(\pi)$
for all $\pi \in \Pi$.
\end{definition}

\begin{definition}[Incentive for Optimal Arrival (I4OA)]
\label{def:I4OA}
A sequential solution concept $\phi$ satisfies \emph{incentive for optimal arrival}
(I4OA) for a TCG $\langle N, v \rangle$ if $\phi_i(\pi^{*}(v),v) \geqslant
\phi_i(\pi,v)$ for all $\pi \in \Pi$ and $i \in N$.
\end{definition}

This property is closely related to I4EA introduced by
\citet{zhangearlyarrivalcooperative}, which requires that agents are not
penalized for arriving earlier in online coalition formation games.  I4OA
strengthens this notion to an order-sensitive setting: rather than merely
protecting early arrivals, it requires that the optimal joining order
constitutes a Nash equilibrium of the arrival game induced by $\phi$.

Note that SE and I4OA serve different normative goals.  SE guarantees the
mechanism is well-defined for every realized sequence, even if agents deviate
from the optimal order or some agents do not join at all.  I4OA is a design
goal that nudges agents toward the optimal order via the reward rule.

In the first part of this paper, we characterize reward-sharing mechanisms
satisfying OIR, I4OA, and SE for basis-satisfiable TCGs.  In the second part,
we define properties analogous to the Shapley properties in classical
cooperative games and characterize the reward-sharing schemes satisfying them.
Finally, we establish the non-existence of sequential solution concepts satisfying both
sets of properties simultaneously.  Detailed proofs are deferred to the
appendix.

\section{The \clAlg{} Class of Mechanisms}
\label{sec:seqsh}

In this section, we introduce a class of reward-sharing mechanisms that we
later show to uniquely satisfy OIR, I4OA, and SE.  We restrict our attention
to basis-satisfiable TCGs (\Cref{def:basis}), and the following lemma
justifies why this restriction is necessary.

\begin{lemma}
\label{lem:I4OA_efficiency_and_basis}
For a TCG \seqgame{}, if there exists a sequential solution concept $\phi$ satisfying
I4OA and SE, then \seqgame{} must be basis-satisfiable.
\end{lemma}
\begin{psketch}
Given a TCG \seqgame{} and a sequential solution concept $\phi$ satisfying I4OA and SE,
setting $x = \phi(\pi^*)$ meets the basis conditions.  Condition~(ii) follows
directly from SE applied to $\pi^*$.  Condition~(i) follows from I4OA, which
bounds $\phi_i(\pi)$ by $\phi_i(\pi^*)$ for all $i$ and $\pi$, and SE
ensures $v(\pi) = \sum_i \phi_i(\pi)$, implying the required inequality.
\end{psketch}

\Cref{lem:I4OA_efficiency_and_basis} shows that basis-satisfiability is
necessary for I4OA and SE to hold together, and \Cref{thm:properties_iff_seqsh}
below shows, constructively, that it is also sufficient. Basis-satisfiability
is therefore a feasibility condition of the problem itself rather than a
restriction of our mechanisms: on a TCG that is not basis-satisfiable, no
sequential solution concept satisfies both I4OA and SE. Our mechanisms are
thus defined exclusively for basis-satisfiable TCGs, and \textsc{SeqShare}
returns null otherwise. Unlike the core of a classical cooperative game, the
basis can be empty even when the TCG is convex (\Cref{app:convex}).

\noindent\textsc{Remark.}  The existence of a basis solution does not imply
that every $\phi$ will automatically satisfy I4OA and SE.  In fact, for any
two of I4OA, OIR, and SE, there exists a sequential solution concept satisfying both
while violating the third, even in games with a non-empty basis.  The
following example shows this for each pair.

\begin{example} 
Consider the TCG: $v(1) = 1, v(12) = 2, v(2) = 1, v(21) = 3$. 
\noindent
\begin{minipage}[t]{0.38\linewidth}
\centering
\small
\captionof{table}{I4OA-SE-OIR independence.}
\label{tab:solutions}
\vspace{0.3em}
\begin{tabular}{|c|c|c|c|c|}
\hline
$\pi$ & $v(\pi)$ & $\phi$ & $\phi'$ & $\phi''$ \\
\hline
1  & 1 & $(0,0)$ & $(1,0)$ & $(1,0)$ \\
2  & 1 & $(0,0)$ & $(0,1)$ & $(0,1)$ \\
12 & 2 & $(0,0)$ & $(0,2)$ & $(2,0)$ \\
21 & 3 & $(1,2)$ & $(1,2)$ & $(1,2)$ \\
\hline
\end{tabular}
\end{minipage}\hfill
\begin{minipage}[t]{0.58\linewidth}
It is easy to check that the game has a basis solution. 
    In this example, $\pi^* = 21$, $x_1=1,x_2=2$. Now, consider the sequential solution concepts below: (i)~$\phi$ satisfies OIR and I4OA but not SE since $v(12)\neq \phi_1(12)+\phi_2(12)$; (ii)~$\phi'$ satisfies SE and I4OA but not OIR as {$\phi'_1(1)=1>0=\phi'_1(12)$}; (iii)~$\phi''$ satisfies SE and OIR but not I4OA since {$\phi''_1(12)=2>1=\phi''_1(21)$}%
. This example also exhibits that no two of these properties imply the third.
\end{minipage}
\end{example}
Our desired class of sequential solution concepts starts with the basis conditions to
check if a solution exists (and returns \textsc{Null} if the basis is empty).
If the basis conditions hold, \Cref{algo:SeqSh} yields a class of solution
concepts satisfying OIR, I4OA, and SE.  It is a class because there are
multiple possible sequential solution concepts $\phi$ given by this algorithm.  However,
we show that each satisfies the three properties, and any sequential solution concept
satisfying all three is in this class.

The class \clAlg{} operates in two phases.  First, an \emph{offline
initialization phase} computes the basis solution via the \textsc{Basis}
subroutine (line~1 of \Cref{algo:SeqSh}), which amounts to solving a linear
program and is therefore solvable in polynomial time in $|\Pi|$
{(\Cref{sec:runtime})}.  This phase %
requires global knowledge of $v$ but is performed offline---once, before any agent arrivals---analogously to how a solution concept is computed before play begins in classical cooperative games.  Second, the \emph{reward
assignment} at each arrival step is online: given the current sequence and
the precomputed upper bounds $x$, \clAlg{} assigns rewards without any
knowledge of future arrivals.

\clAlg{} performs the following steps:
\begin{enumerate}[leftmargin=14pt, itemsep=0pt, parsep=0pt, topsep=0pt]
  \item It checks for a basis solution.  The returned value $x_i$ is the
    upper bound on agent $i$'s reward for all sequences, so $\phi_i(\pi^*)$
    is set to $x_i$.
  \item For a sequence $\pi$ with $\pi(1)=i$, the algorithm sets $\phi_i(i)=v(i)$, which becomes the lower bound for $\phi_i(\pi)$ for every sequence having $i$ as a prefix.
  \item It computes the marginal contribution of $\pi(2)=j$ and divides it among $i$ and $j$ so that $\phi_i(ij)$ and $\phi_j(ij)$ lie within the upper and lower bounds.  These bounds are then updated for every sequence containing $ij$ as a prefix.
  \item The algorithm iterates, assigning rewards within the bounds at each arrival and updating the lower bound to the most recently assigned value.
\end{enumerate}

\begin{figure}[h!]
    \centering
    \resizebox{0.24\columnwidth}{!}{
    \begin{subfigure}[b]{0.3\textwidth}
        \centering
        \begin{tikzpicture}[]
            \draw[fill=orange!20] (0,0) rectangle (0.5,2);
            \draw[fill=orange!20] (1.0,0) rectangle (1.5,3);
            \draw[fill=orange!20] (2.0,0) rectangle (2.5,1);
            \draw[fill=orange!20] (3.0,0) rectangle (3.5,2.5);
            \draw[fill=orange!50] (0,0) rectangle (0.5,0.5);
            \node at (0.25,-0.5) {\textcolor{teal}{P1}};
            \node at (1.25, -0.5) {P2};
            \node at (2.25, -0.5) {P3};
            \node at (3.25, -0.5) {P4};
             \draw (3.75,3) -- (3.75,-0.5);
            \draw[fill=orange!20] (0.5,3.5) rectangle (0.9,3.9);
           \node[anchor=west] at (0.9,3.7) {\footnotesize{Possible additional reward}};
        \end{tikzpicture}
    \end{subfigure}}
     \resizebox{0.24\columnwidth}{!}{
    \begin{subfigure}[b]{0.3\textwidth}
        \centering
        \begin{tikzpicture}[]
            \draw[fill=orange!20] (0,0) rectangle (0.5,2);
            \draw[fill=orange!20] (1.0,0) rectangle (1.5,3);
            \draw[fill=orange!20] (2.0,0) rectangle (2.5,1);
            \draw[fill=orange!20] (3.0,0) rectangle (3.5,2.5);
            \draw[fill=orange!90] (0,0) rectangle (0.5,0.5);
            \draw[fill=orange!50] (0,0.5) rectangle (0.5,0.75);
            \draw[fill=orange!50] (1,0) rectangle (1.5,0.7);
            \draw[draw=none, fill=none] (4,0) rectangle (4,2);
            \node at (0.25,-0.5) {P1};
            \node at (1.25, -0.5) {\textcolor{teal}{P2}};
            \node at (2.25, -0.5) {P3};
            \node at (3.25, -0.5) {P4};
             \draw (3.75,3) -- (3.75,-0.5);
             \draw[fill=orange!50] (1.4,3.5) rectangle (1.8,3.9);
           \node[anchor=west] at (1.8,3.7) {\footnotesize{Marginal Contribution of new player}};
        \end{tikzpicture}
    \end{subfigure}}
     \resizebox{0.24\columnwidth}{!}{
    \begin{subfigure}[b]{0.3\textwidth}
        \centering
        \begin{tikzpicture}[]
            \draw[fill=orange!20] (0,0) rectangle (0.5,2);
            \draw[fill=orange!20] (1.0,0) rectangle (1.5,3);
            \draw[fill=orange!20] (2.0,0) rectangle (2.5,1);
            \draw[fill=orange!20] (3.0,0) rectangle (3.5,2.5);
            \draw[fill=orange!90] (0,0) rectangle (0.5,0.75);
            \draw[fill=orange!90] (1,0) rectangle (1.5,0.7);
            \draw[fill=orange!50] (2,0) rectangle (2.5,1);
            \draw[fill=orange!50] (0,0.75) rectangle (0.5,1);
            \draw[fill=orange!50] (1,0.7) rectangle (1.5,1.5);
            \draw[draw=none, fill=none] (4,0) rectangle (4,2);
            \node at (0.25,-0.5) {P1};
            \node at (1.25, -0.5) {P2};
            \node at (2.25, -0.5) {\textcolor{teal}{P3}};
            \node at (3.25, -0.5) {P4};
             \draw (3.75,3) -- (3.75,-0.5);
        \end{tikzpicture}
    \end{subfigure}}
     \resizebox{0.24\columnwidth}{!}{
    \begin{subfigure}[b]{0.3\textwidth}
        \centering
         \begin{tikzpicture}[]
            \draw[fill=orange!20] (0,0) rectangle (0.5,2);
            \draw[fill=orange!20] (1.0,0) rectangle (1.5,3);
            \draw[fill=orange!20] (2.0,0) rectangle (2.5,1);
            \draw[fill=orange!20] (3.0,0) rectangle (3.5,2.5);
            \draw[fill=orange!90] (0,0) rectangle (0.5,1);
            \draw[fill=orange!90] (1,0) rectangle (1.5,1.3);
            \draw[fill=orange!90] (2,0) rectangle (2.5,1);
             \draw[fill=orange!50] (3,0) rectangle (3.5,0.8);
             \draw[fill=orange!50] (0,1) rectangle (0.5,1.5);
             \draw[fill=orange!50] (1,1.3) rectangle (1.5,1.8);
             \draw[draw=none, fill=none] (4,0) rectangle (4,2);
            \node at (0.25,-0.5) {P1};
            \node at (1.25, -0.5) {P2};
            \node at (2.25, -0.5) {P3};
            \node at (3.25, -0.5) {\textcolor{teal}{P4}};
             \draw[fill=orange!90] (0.5,3.5) rectangle (0.9,3.9);
            \node[anchor=west] at (0.9,3.7) {\footnotesize{Reward already assigned}};
        \end{tikzpicture}
    \end{subfigure}}
    \caption{(a) The first player arrives, and all the value generated is assigned to player P1. (b) The second player P2 arrives, and the marginal value generated by the arrival of this player is distributed among P1 and P2 in a way that ensures the total reward of each agent is less than the upper bound. The total reward for any agent is the sum of the reward in the previous round and the share of the marginal increase assigned in the current round. (c) The third player P3 arrives, and the marginal value generated by P3 is distributed among P1, P2, and P3, keeping total rewards lower than the upper bound. (d) Finally, P4 arrives, and the marginal value generated by this player can only be divided among P1, P2, and P4, as P3 has already reached its maximum possible reward.}
    \label{fig:buckets}
\end{figure}
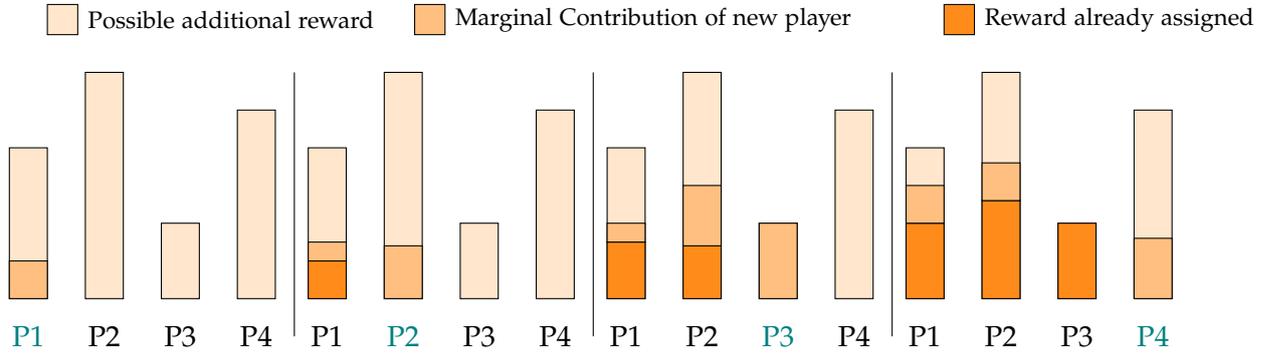

\Cref{fig:buckets} illustrates one particular run of \clAlg{} on a
four-player game; it is an example of how rewards fill the agents' caps.  A complete numerical trace of \Cref{algo:SeqSh} on a three-agent TCG, listing the vector returned by \basis{} and the vector $y$ returned by \impr{} at every arrival step, is given in \Cref{app:walkthrough}. The non-uniqueness of the sequential solution concept arises from two sources: (i)~the basis solution may not be unique, and (ii)~the \impr{} function returns a non-unique $y$ vector.  
\begin{algorithm}[h]
\caption{\clAlg{} Class of Mechanisms}
\label{algo:SeqSh}
\small
\begin{minipage}[t]{0.54\linewidth}
\begin{algorithmic}[1]
\Require TCG $\langle N,v \rangle$
\State Find $\pi^*(v) = \argmax_\pi v(\pi)$, $x = \basis(N,v,\pi^*(v))$
\If {$x$ = \nl{}}
  Output \nl{}, \textsc{exit}
\Else \ \ $\phi(\pi^*(v)) = x$; $\phi(\pi) = \mathbf{0}, \ \forall \pi \neq \pi^*(v)$
\EndIf
\For{$i \in N$}\ $\phi_i(\{i\}) = v(\{i\})$ \EndFor
\For{$k = 2,\ldots,n$}
  \For{$\pi \in \Pi,\ |\pi|=k-1$}
    \For{$j \in N \setminus P(\pi)$}
      \State $y = \impr(j,\pi,\phi(\pi),\phi(\pi^*(v)))$
      \State $\phi_i(\pi+j) = \phi_i(\pi)+y_{i},\ \forall i \in P(\pi)$
      \State $\phi_j(\pi+j) = y_{j}$
    \EndFor
  \EndFor
\EndFor
\State \textbf{Output:} $\phi$
\end{algorithmic}
\end{minipage}\hfill
\begin{minipage}[t]{0.43\linewidth}
\begin{algorithmic}[1]
\Procedure{$\basis(N,v,\pi^*(v))$}{}
  \If{solution to \Cref{eq:basis} exists}
    \State \textbf{return} $x$ of \Cref{eq:basis} w.r.t.\ $\pi^*(v)$
  \Else\ \textbf{return} \nl{}
  \EndIf
\EndProcedure
\end{algorithmic}
\vspace{0.5em}
\begin{algorithmic}[1]
\Procedure{$\impr(j,\pi,\phi(\pi),\phi(\pi^*))$}{}
  \State $y_{i} \in [0,\phi_i(\pi^*)-\phi_i(\pi)],\ \forall i \in P(\pi+j)$
  \State $\sum_{i \in P(\pi+j)} y_{i} = v(\pi+j)-v(\pi)$
  \State \textbf{return} $y$
\EndProcedure
\end{algorithmic}
\end{minipage}
\end{algorithm}

\noindent\textsc{Remark.}  The existence of a basis solution is a property
of the game \seqgame{}, not of a sequential solution concept.

\subsection{Properties of \clAlg{}}

\Cref{lem:I4OA_efficiency_and_basis} states that a basis solution is
necessary to satisfy I4OA and SE.  The following theorem shows that it is
also sufficient, and that any solution satisfying all three properties belongs
to \clAlg{}.

\begin{theorem}
\label{thm:properties_iff_seqsh}
Given a basis-satisfiable TCG \seqgame{}, a {nonnegative} sequential solution concept %
$\phi$ satisfies OIR, I4OA, and SE for \seqgame{} if and only if it belongs to the class
\clAlg{}.
\end{theorem}

\begin{psketch}
If \seqgame{} violates the basis conditions, then by
\Cref{lem:I4OA_efficiency_and_basis} no sequential solution concept can satisfy I4OA
and SE, and \clAlg{} returns \nl{}.  Hence, assume the basis conditions hold.

We first prove the reverse direction constructively using \Cref{algo:SeqSh}.
The algorithm initializes the allocation using \basis{} on $\pi^*(v)$ and
assigns zero elsewhere.  The construction proceeds by induction on sequence
length.

For length-one sequences, SE holds trivially, I4OA follows from the basis
conditions, and OIR holds by monotonicity.  Assume the properties hold for
all sequences of length $k-1$.  For a sequence $\pi+i$ of length $k$, the
function \impr{} allocates the marginal reward so that payoffs are
non-decreasing (OIR), bounded above by $\phi(\pi^*)$ (I4OA), and sum to
$v(\pi+i)$ (SE).

For the forward direction, suppose $\phi$ satisfies I4OA, SE, and OIR but
$\phi \notin \clAlg{}$.  For all $\phi' \in \clAlg{}$, let $\pi'$ be the
shortest sequence where $\phi'$ and $\phi$ differ; among these, choose the
$\phi'$ corresponding to the longest such sequence.  Call this sequence
$\pi_{\textup{longshort}}$.  Using the properties of $\phi$, we construct a
solution in \clAlg{} that matches $\phi$ on all sequences up to and including
$|\pi_{\textup{longshort}}|$, contradicting the choice of
$\pi_{\textup{longshort}}$.  Hence, every solution satisfying I4OA, SE, and
OIR belongs to \clAlg{}.
\end{psketch}

{
\subsection{Running Time of \clAlg{}}
\label{sec:runtime}

The input to \clAlg{} is the full table of worths $\{v(\pi)\}_{\pi \in \Pi}$, so its natural size measure is $|\Pi|$.  Counting sequences by their length $k$, there are $n!/(n-k)!$ sequences of length $k$, and therefore

\begin{equation}
\label{eq:Pi-size}
  |\Pi| \;=\; \sum_{k=1}^{n} \frac{n!}{(n-k)!} \;=\; \lfloor e \cdot n! \rfloor - 1 .
\end{equation}
For example, $|\Pi| = 15$ when $n=3$ and $|\Pi| = 64$ when $n=4$.

\begin{prop}
\label{prop:runtime}
For every TCG \seqgame{}, \Cref{algo:SeqSh} can be implemented to run in
time $T_{\mathrm{LP}}(n, |\Pi|) + O(n\,|\Pi|)$, where $T_{\mathrm{LP}}(n,m)$ is
the time to solve a linear feasibility problem with $n$ variables and $m$
constraints.  In particular, the running time is polynomial in $|\Pi|$.  Once
the basis vector $x$ is computed, the reward of every agent at each arrival
step can be assigned online in $O(n)$ time.
\end{prop}

\begin{psketch}
The \basis{} LP has $n$ variables and $|\Pi|+1$ constraints, and the loop
calls \impr{} once per sequence of length at least two, each call being a
greedy $O(n)$ filling of slacks that never fails by the basis conditions
(\Cref{app:runtime}).
\end{psketch}

Since $|\Pi| \geqslant n!$, the running time is exponential in the number of
agents $n$.  This is the same barrier faced by the exact computation of the
classical Shapley value, which is \#P-complete for compactly represented games such as weighted voting games~\citep{deng1994complexity}, and it is unavoidable for any method
that reads the entire worth table.  As with the Shapley value, we view the
axiomatic foundations as the first step; tractable special classes of TCGs
and sampling-based approximations in the spirit
of~\citet{castro2009polynomial} are natural directions, which we discuss in
\Cref{sec:conclusion}.
}
\section{Properties Inspired by Shapley in TCGs}
\label{sec:shapley-prop}

In classical cooperative games, the characteristic function $v$ depends on a coalition, whereas in TCGs, it depends on a player sequence. To compare solution concepts, we distinguish between \emph{sequential solution concepts} $\phi \in \Phi = \{\phi : \Pi \times V \to \mathbb{R}^n\}$, which depend on a sequence $\pi$, and \emph{extended solution concepts} $\psi \in \Psi = \{\psi : V \to \mathbb{R}^n\}$, which do not, where $V$ is the set of monotone characteristic functions. %
Intuitively, $\phi_i(\pi)$ represents agent $i$'s payoff given a specific formation order $\pi$ (e.g., Hospital~$A$'s reward when joining in order $\pi$), whereas $\psi_i$ provides a single, sequence-independent payoff for agent $i$. 
Any sequential solution concept $\phi$ can be \emph{reduced} to an extended
solution concept $\overline{\phi}$ by averaging $\phi(\pi)$ over all full
sequences, as made precise in \Cref{def:reduction_of_sol}.

\begin{definition}
\label{def:reduction_of_sol}
Given a TCG \seqgame{} and $\phi \in \Phi$, the reduction of $\phi$ to
$\overline{\phi} \in \Psi$ is defined as $\overline{\phi}_i=
\frac{1}{n!}\sum_{\pi \in \Pi : P(\pi) = N}\phi_i(\pi)$.
\end{definition}

The map $\phi \mapsto \overline{\phi}$ is many-to-one and has no meaningful
inverse. An extended solution concept assigns exactly one reward per agent,
independent of the realized order, so it cannot recover the order-specific
information carried by a sequential solution concept. None of our results
requires a reverse construction, since the incompatibility in
\Cref{sec:bestofboth} only asks whether the guarantees of \clAlg{} survive the forward reduction.

In parallel with the Shapley properties, we now define additivity,
efficiency, and null player for both types of solution concepts.  Properties
for the sequential solution concepts are called \emph{sequential}
properties; those for extended solution concepts are called \emph{extended}
properties.  Sequential efficiency is already defined in \Cref{def:SE}.

\begin{definition}[Sequential Additivity (SA)]
A sequential solution concept $\phi \in \Phi$ is \emph{sequentially additive} if for
every pair of TCGs $\langle N,v\rangle$ and $\langle N,u \rangle$,
$\phi_i(\pi,u)+\phi_i(\pi,v)=\phi_i(\pi,u+v)$ for all $\pi \in \Pi$ and
$i \in N$.
\end{definition}

\begin{definition}[Sequential Null Player (SNP)]
A sequential solution concept $\phi \in \Phi$ satisfies \emph{sequential null player}
if for every TCG \seqgame{},
$v(\pi+i)=v(\pi)\ \forall \pi \in \Pi_{-i}\ \implies\ \phi_i(\pi)=0\ \forall
\pi \in \Pi$.
\end{definition}

\begin{definition}[Extended Efficiency (EE)]
\label{def:ext_eff}
An extended solution concept $\psi \in \Psi$ satisfies \emph{extended
efficiency} if for every TCG \seqgame{},
$\sum_{i \in N} \psi_i(v)=\frac{1}{n!}\sum_{\pi \in \Pi : P(\pi) = N}
v(\pi)$.
\end{definition}

\begin{definition}[Extended Null Player (ENP)]
\label{def:ext_null}
An extended solution concept $\psi \in \Psi$ satisfies \emph{extended null
player} if for every TCG \seqgame{},
$v(\pi+i)=v(\pi)\ \forall \pi \in \Pi_{-i}\ \implies\ \psi_i(v)=0$.
\end{definition}

This definition of extended null player is stronger than that in
\citet{sanchez1997values}, where a player is null only if inserting it at any
position of a sequence does not improve worth.  Our null player only requires
the agent to be added at the end.  Since the antecedent of our definition is
weaker, ENP is a stronger property.  Their characterization requires four
properties, including symmetry, while we need only three
(\Cref{sec:Shapley-solutions}).  We show in \Cref{example:sanchez-not-same}
that even using their definitions of the Shapley properties, the
TCG-appropriate properties conflict with the Shapley-inspired ones.

\begin{example}
\label{example:sanchez-not-same}
Consider the TCG with $v(1)=1, v(12)=1, v(2)=1, v(21)=2$.  The only
solution in \clAlg{} is $\phi(1)=(1,0), \phi(2)=(0,1), \phi(12)=(1,0),
\phi(21)=(1,1)$, with reduction $\overline{\phi}=(1,0.5)$.  The solution of
\citet{sanchez1997values} for this game is $\psi=(0.75,0.75)$.  Hence there
exist games for which no $\phi \in \clAlg{}$ can be reduced to the solution
of \citet{sanchez1997values}.
\end{example}

\begin{definition}[Extended Additivity (EA)]
\label{def:ext_additivity}
An extended solution concept $\psi \in \Psi$ satisfies \emph{extended
additivity} if for every pair of TCGs $\langle N,v\rangle$ and $\langle
N,u\rangle$, $\psi_i(u)+\psi_i(v)=\psi_i(u+v)$ for all $i \in N$.
\end{definition}

One important property for the Shapley value is \emph{symmetry}.  A
\emph{swap} of agents $i$ and $j$ in sequence $\pi$ is defined as: (a)~$i$
replaced by $j$ if only $i \in P(\pi)$, (b)~$j$ replaced by $i$ if only
$j \in P(\pi)$, or (c)~both positions swapped if $i,j \in P(\pi)$.  This is
denoted by $\pi_{\swap{i}{j}}$.

\begin{definition}[Sequential Symmetry (SS)]
\label{def:seq_symmetry}
A sequential solution concept $\phi \in \Phi$ satisfies \emph{sequential symmetry} if
for every TCG \seqgame{},
$v(\pi) = v(\pi_{\swap{i}{j}})\ \forall \pi \in \Pi \implies \phi_i(\pi)
= \phi_j(\pi_{\swap{i}{j}})\ \forall \pi \in \Pi$.
\end{definition}

\begin{definition}[Extended Symmetry (ES)]
\label{def:ext_symmetry}
An extended solution concept $\psi \in \Psi$ satisfies \emph{extended
symmetry} if for every TCG \seqgame{},
$v(\pi)=v(\pi_{\swap{i}{j}})\ \forall \pi \in \Pi\ \implies\ \psi_i=\psi_j$.
\end{definition}

However, we will see in \Cref{sec:symmetry-not-necessary} that each of these
symmetry properties is a consequence of the other three properties in the
respective solution spaces.

\section{Solutions Satisfying Shapley-inspired Properties}
\label{sec:Shapley-solutions}

In this section, we introduce two solution concepts, one in $\Phi$ and one in
$\Psi$, and show that each uniquely satisfies a corresponding set of
properties.

We define the sequential solution concept $\MargSol \in \Phi$ inspired by marginal
contributions in classical cooperative games:
\begin{equation}
\label{eq:margsol}
  \MargSol_i(\pi)=v(\pi_i+i)-v(\pi_i), \quad \forall \pi \in \Pi,\ i \in N.
\end{equation}

Define the \emph{Extended Shapley Value} $\ExtSh$ by:
\begin{equation}
\label{eq:ExtShapley}
  \ExtSh_i = \frac{1}{n!}\sum_{\pi \in \Pi:|\pi| = n} \MargSol_i(\pi).
\end{equation}

Observe that $\ExtSh = \overline{\MargSol}$, i.e., \ExtSh{} is the reduced
version of \MargSol{}.  The following lemma shows that sequential properties
are stronger than their extended counterparts.

\begin{lemma}
\label{lemma:reduced-property-satisfaction}
For every TCG \seqgame{} and every $\phi \in \Phi$: (1)~$\phi$ satisfies SE
$\implies \overline{\phi}$ satisfies EE; (2)~$\phi$ satisfies SA $\implies
\overline{\phi}$ satisfies EA; (3)~$\phi$ satisfies SNP $\implies
\overline{\phi}$ satisfies ENP.
\end{lemma}
\begin{psketch}
Averaging the sequential property over all full-length sequences gives the
corresponding extended property.
\end{psketch}

Our next goal is to show that the three properties are uniquely satisfied by
\MargSol{}.  We use a construction analogous to the Shapley uniqueness
proof~\citep[Chapter~19]{maschler2020game}.  We define the \emph{carrier
games} in the context of TCGs as follows.

\begin{definition}[Carrier Game]
\label{def:carrier}
For any sequence $\pi \in \Pi$, the carrier game over $\pi$ is the simple
game\footnote{Following classical cooperative game terminology: a simple game
is a TCG where the worth of any sequence is either zero or one.}
$\langle N,u_{\pi} \rangle$ where $u_{\pi}(\pi') = 1$ if $\pi \sqsubset
\pi'$ and $0$ otherwise.
\end{definition}

\begin{lemma}
\label{lemma:carrier_games_basis}
Every TCG $\langle N, v \rangle$ is a linear combination of carrier games.
\end{lemma}
\begin{psketch}
The set of all carrier games forms a linearly independent set of size $|\Pi|$,
forming a spanning set for the vector space of all TCGs.
\end{psketch}

\subsection{Uniqueness of \MargSol{}}
\label{sec:unique-margsol}

\begin{lemma}
\label{lemma:margsol_carrier}
Consider a TCG $\langle N,u_{\pi,\alpha} \rangle$ where $u_{\pi,\alpha}(\pi')
= \alpha$ if $\pi \sqsubset \pi'$ and $0$ otherwise.  If $\phi \in \Phi$
satisfies SE and SNP, then $\phi_i(\pi',u_{\pi,\alpha}) = \alpha$ if
$i = \ell(\pi)$ and $\pi \sqsubset \pi'$, and $0$ otherwise.
\end{lemma}

\begin{psketch}
Intuitively, all players except $\ell(\pi)$ are null players in $\langle
N,u_{\pi,\alpha} \rangle$, so by efficiency the last player must receive the
entire reward $\alpha$.
\end{psketch}

\begin{theorem}
\label{Thm:MargSol_unique}
For every TCG \seqgame{}, a sequential solution concept $\phi \in \Phi$ satisfies SE,
SA, and SNP if and only if $\phi \equiv \MargSol$.
\end{theorem}
\begin{psketch}
($\Leftarrow$) If player $i$ is a sequential null player, then
$\MargSol_i(\pi) = 0$ for all $\pi$.  Additivity (SA) holds by linearity of
marginal contributions, and efficiency (SE) follows by telescoping the sum
of incremental values across the sequence.

($\Rightarrow$) Any TCG can be written as a sum of carrier games.  For each
carrier game, \MargSol{} is the unique solution satisfying SE and SNP.  If
another solution $\phi$ also satisfies SE, SA, and SNP, it must agree with
\MargSol{} on all carrier games.  By linearity and the basis decomposition,
$\phi$ must equal \MargSol{} on the entire game.
\end{psketch}

Although \ExtSh{} is the averaged version of \MargSol{} and \MargSol{} uniquely satisfies SE, SA, and SNP (\Cref{Thm:MargSol_unique}), this does not automatically imply that \ExtSh{}
uniquely satisfies EE, EA, and ENP. \citet[Theorem~1]{NOWAK1994150} established that \ExtSh{} indeed uniquely satisfies EE, EA, and ENP.  We restate this theorem and present an alternative proof in \Cref{sec:unique-extsh}.

\noindent\textsc{Remark.} 
Unlike the classical Shapley value, \MargSol{} and \ExtSh{} are uniquely characterized by three axioms rather than four: symmetry is redundant because in a TCG carrier game only the last player is non-null, and receives the entire worth, so symmetry is not
required. We show in \Cref{sec:symmetry-not-necessary} that symmetry follows from the other three properties.

\section{\clAlg{} and Shapley-compliant: Best of Both Worlds}
\label{sec:bestofboth}

The \clAlg{} class characterizes three desirable properties for TCGs: OIR,
SE, and I4OA, embodying the spirit of \emph{sequential incentives}---rewards
are shaped by the realized arrival order, protecting early participants and
aligning individual incentives with the socially optimal sequence.  \ExtSh{},
on the other hand, uniquely characterizes three Shapley-equivalent properties:
EA, EE, and ENP, yielding Shapley-based allocations where each agent receives
a reward reflecting their incremental worth averaged over all possible
sequences.  A solution concept satisfying both sets of properties
simultaneously would be the ideal candidate.  The central result of this
section shows that these two principles carve out provably disjoint regions
of the solution space, and that this holds even for convex and simple TCGs.

To show the result for general basis-satisfiable TCGs, we first make the
following observation.

\begin{lemma}
\label{lemma:extended_space_proj}
For a TCG \seqgame{} and $\phi \in \clAlg{}$, if $\overline{\phi}=\ExtSh{}$
then for all $i \in N$:
\begin{equation}
\label{eq:necessary_extSh_SeqSh}
  \phi_i(\pi^*(v)) \geqslant \frac{1}{n!}\sum_{\pi \in \Pi : P(\pi) = N}
  \bigl(v(\pi_i+i)-v(\pi_i)\bigr) = \ExtSh_i.
\end{equation}
\end{lemma}
\begin{psketch}
Suppose $\phi_i(\pi^*(v))$ is strictly less than the average marginal
contribution.  By I4OA, $\phi_i(\pi) \leq \phi_i(\pi^*(v))$ for all $\pi$,
so the average $\overline{\phi}_i$ is also strictly less than $\ExtSh_i$,
contradicting $\overline{\phi} = \ExtSh$.
\end{psketch}

Hence, \Cref{eq:necessary_extSh_SeqSh} is necessary for a solution in
\clAlg{} to reduce to \ExtSh{}.  The following example shows a game that
violates this condition.

\begin{example}
 \label{example:no-general-reduction}
Consider the TCG, $v(1) = 1, v(12) = 3, v(123) = a , v(13) = 4, v(132) = 6  ,
v(2) = b, v(21) = 4, v(213) = 8  , v(23) = 5, v(231) = 7  ,
v(3) = 4, v(31) = 5, v(312) = a  , v(32) = 5, v(321) = 7$ with $a=7,b=3$. Clearly, $v(\pi^*) = 8$. The basis solution (\Cref{eq:basis}) dictates that $x_1 \geqslant v(1)=1, x_2 \geqslant v(2)=3, x_3 \geqslant v(3)=4$ and $x_1+x_2+x_3=8$, which implies that the basis solution must be $x_1=1, x_2=3, x_3=4$. This is the only possible basis solution and therefore any $\phi \in \clAlg{}$ assigns $\phi_i(\pi^*)=x_i$. Routine calculations for Player 1 {give} %
 $\ExtSh{}_1=\frac{8}{6}$. However, $\phi_1(\pi^*) = 1 < \frac{8}{6} = \ExtSh{}_1$, which violates \Cref{eq:necessary_extSh_SeqSh}. 
\end{example}

\noindent\textsc{Remark.}  Since no solution in \clAlg{} reduces to \ExtSh{},
no solution in \clAlg{} can equal \MargSol{}. \MargSol{} satisfies SE and OIR (marginal contributions are non-negative for monotone games) but violates I4OA, confirming that no sequential solution concept can satisfy both sets of properties simultaneously.

Having established that Shapley-inspired properties and I4OA define provably
disjoint axiomatic worlds for general TCGs, we now ask whether additional
structure on the game can bring these two worlds closer. In \Cref{app:convex,app:simple} we consider convex and simple {TCGs, respectively,} %
 and conclude that even in these restricted classes the barrier remains.

\section{Conclusions and Future Work}
\label{sec:conclusion}
Many AI systems involve agents contributing to a shared objective
sequentially, where the order of participation shapes the outcome.
This paper develops the axiomatic foundations for reward sharing in such
settings.
We characterized all reward-sharing mechanisms satisfying three natural
sequential properties via the \clAlg{} class, and
showed that a Shapley-analog uniquely satisfies the order-sensitive
extensions of efficiency, additivity, and null player.
Our central result is a sharp impossibility: no mechanism can
simultaneously satisfy sequential incentive properties and Shapley-based fairness, even for convex and simple TCGs.

This paper has the following assumptions and limitations.
\emph{(i) Complete information.} The worth function $v$ is known to all before anyone commits. \emph{(ii) Basis-satisfiability.} The positive characterization (\Cref{thm:properties_iff_seqsh}) applies only to basis-satisfiable TCGs.  By \Cref{lem:I4OA_efficiency_and_basis} this is not an artifact of \clAlg{}, since no sequential solution concept can satisfy I4OA and SE on other games; but it does mean that in such games a designer must give up one of the two properties, and, as \Cref{app:convex} shows, even convexity does not rule this case out. \emph{(iii) Monotone, nonnegative worths and nonnegative rewards.} We assume $v$ is monotone along prefixes and nonnegative, and we restrict attention to nonnegative sequential solution concepts.  Games in which a late arrival can destroy value, or mechanisms that charge agents, are outside our scope. \emph{(iv) Computation.} \clAlg{} runs in time polynomial in $|\Pi|$ but exponential in $n$ (\Cref{prop:runtime}), so it suits a small consortium, such as a few hospitals, rather than large multi-agent systems; the paper is purely theoretical and contains no empirical evaluation.

Open directions are to single out a unique mechanism within \clAlg{}, to find
subclasses of TCGs where sequential and Shapley-like axioms co-exist or where
\clAlg{} is computable in time polynomial in $n$, and to design sampling-based
approximations in the spirit of~\citet{castro2009polynomial}.

\begin{ack}
{SN acknowledges support from an Amazon Research Grant (2D-09926960) and two
grants from the Anusandhan National Research Foundation (ANRF), Government of
India (CRG/2023/001442 and ANRF/ARG/2025/008746/QSS).  DD acknowledges support
from the Asha Navani Fellowship.}

The authors declare no competing interests.
\end{ack}

\bibliographystyle{plainnat}
\bibliography{references}

\appendix

\section{Appendix to \Cref{sec:seqsh}}

\subsection{Proof of \Cref{lem:I4OA_efficiency_and_basis}}
\begin{proof}
    Consider a TCG \seqgame{} and a sequential solution concept $\phi$ that satisfies I4OA and SE in the game. We claim that $x = \phi(\pi^*)$ satisfies the basis condition.
    \begin{itemize}
        \item Condition~(ii) of \Cref{eq:basis} is satisfied by applying SE of $\phi$ for $\pi^*$.
        \item Since $\phi$ also satisfies I4OA, $\phi_i(\pi) \leqslant \phi_i(\pi^*), \forall \pi \in \Pi, \forall i \in N$. Hence, $\sum_{i \in P(\pi)} \phi_i(\pi) \leqslant \sum_{i \in P(\pi)} \phi_i(\pi^*)$. By SE of $\phi$, $v(\pi) = \sum_{i \in P(\pi)} \phi_i(\pi)$. Therefore, we get $\sum_{i \in P(\pi)} \phi_i(\pi^*) \geqslant v(\pi)$, which is condition~(i) of \Cref{eq:basis}.
    \end{itemize}
    Hence $\phi(\pi^*)$ satisfies basis conditions.
\end{proof}

\subsection{Proof of \Cref{thm:properties_iff_seqsh}}
\begin{proof}
If the game \seqgame{} does not satisfy the basis conditions, then from \Cref{lem:I4OA_efficiency_and_basis} we know that no $\phi$ can satisfy I4OA and SE. In that case, \clAlg{} returns \nl{} and hence the theorem holds. For the rest of the proof, we will assume that the game satisfies the basis conditions.

First, we prove the reverse direction, constructively via \Cref{algo:SeqSh}. 
The function \basis{} returns a valid {\em partial} allocation $x$, which is assigned to the value of the sequential solution concept for the specific sequence $\pi^*(v)$, i.e., $\phi(\pi^*(v))$, and $\phi$ is initialized to zero for all other sequences%
{. The rest of the proof is by induction on the length of the sequence $\pi$, with the singleton sequences as the base case. Since the empty sequence is not an element of $\Pi$, no statement about it is needed.
    
    \emph{Base case.} Consider a singleton sequence $(i)$. \Cref{algo:SeqSh} sets $\phi_i(i) = v(i)$, so SE holds for $(i)$ by construction. Property I4OA holds because condition~(i) of the basis conditions (\Cref{eq:basis}), applied to the sequence $(i)$, gives $x_i \geqslant v(i)$, and hence $\phi_i(i) = v(i) \leqslant x_i = \phi_i(\pi^*(v))$. Nonnegativity holds because we assumed $v(i) \geqslant 0$. Finally, OIR places no constraint on a singleton taken by itself, because a singleton has no proper prefix in $\Pi$; its role as a lower bound for longer sequences is handled in the inductive step.}
    
    Hence, by the induction hypothesis, assume that the forward implication of the theorem is true for all $\pi$ such that $|\pi| = k - 1, k \geqslant 2$. We show that the implication is true for every $\pi+i$ of length $k$ as well. Consider the function \impr{}. For every $\pi$ of length $k - 1$ and the existing partial share $\phi(\pi)$, \impr{} allocates the reward to agents in such a way that it is at least $\phi_i(\pi)$ (required by OIR) and at most $\phi_i(\pi^*)$ (required by I4OA)%
{; the newly arrived agent receives $y_i \geqslant 0$, so the rule stays nonnegative. It remains to check that \impr{} always has a feasible output, that is, that the total available slack covers the marginal worth. Summing the slacks over $P(\pi+i)$ and using $\phi_i(\pi) = 0$ for the agent $i \notin P(\pi)$, we obtain
    \[
      \sum_{j \in P(\pi+i)} \bigl(\phi_j(\pi^*) - \phi_j(\pi)\bigr)
      = \sum_{j \in P(\pi+i)} x_j \;-\; \sum_{j \in P(\pi)} \phi_j(\pi)
      \;\geqslant\; v(\pi+i) - v(\pi).
    \]
    Here the inequality uses two facts. The first term is at least $v(\pi+i)$ by condition~(i) of the basis conditions applied to the sequence $\pi+i$, and the second term equals $v(\pi)$ by SE for $\pi$, which holds by the induction hypothesis. Moreover, each individual slack is nonnegative by the induction hypothesis (I4OA for $\pi$), and the marginal $v(\pi+i) - v(\pi)$ is nonnegative because $v$ is monotone; this is where monotonicity is used. Hence a vector $y$ with $0 \leqslant y_j \leqslant \phi_j(\pi^*) - \phi_j(\pi)$ and the required sum always exists.} Therefore, I4OA and OIR are satisfied for each $i \in N$. Also, the construction of the $y$ vectors (reward shares) is such that the sum of rewards of all agents $j \in \pi + i$ equals $v(\pi+i)$, satisfying SE. 

    Next, we prove the forward direction.
    We prove this via contradiction. Suppose there is a sequential solution concept $\phi$ for a given game \seqgame{} that satisfies I4OA, SE, and OIR but is not from the class \clAlg{}. Then we will construct a sequential solution concept $\Tilde{\phi} \in \clAlg{}$ which matches $\phi$ for every sequence and for every player, leading to the contradiction that $\phi \in \clAlg{}$.
    
    For every solution $\phi'$ from class \clAlg{}, there is a shortest length sequence $\pi \in \Pi$ such that $\phi'_i(\pi) \neq \phi_i(\pi)$ for some $i \in P(\pi)$ while $\phi'_i(\pi') = \phi_i(\pi')$ for all $i \in P(\pi')$ and $\pi' \in \Pi: |\pi'| < |\pi|$, i.e., $\pi$ has the minimum length where the sequential solution concepts $\phi$ and $\phi'$ differ.
    
    Choose the $\phi'$ where the length of the sequence $\pi$ is the longest over all $\phi'$s (breaking ties arbitrarily). Call this longest (over sequential solution concepts) shortest (over sequences) sequence to be $\pi_\textup{longshort}$. Note that, by assumption, no  $\phi'' \in \clAlg{}$ can satisfy $\phi''_i(\pi) = \phi_i(\pi)$ for all $i \in P(\pi)$ where $\pi \in \Pi$ and $|\pi| \leqslant |\pi_\textup{longshort}|$. We will construct a $\Tilde{\phi}$ that achieves this and belongs to \clAlg{} to show a contradiction.

    Start with a game where the required properties are satisfiable. Since this game has a sequential solution concept $\phi$ that satisfies I4OA and SE, \basis{} must be non-empty (\Cref{lem:I4OA_efficiency_and_basis}).
    Also, $\phi(\pi^*)$ will be a solution to the basis by \Cref{lem:I4OA_efficiency_and_basis}. Hence, set $\Tilde{\phi}$ such that $\Tilde{\phi}(\pi^*) = \phi(\pi^*)$.

    Construct $\Tilde{\phi} \in \clAlg{}$ as follows: $\Tilde{\phi}_i(\pi') = \phi_i(\pi'), \forall i \in P(\pi'), \forall \pi':|\pi'| < |\pi_\textup{longshort}|$. Clearly, this is feasible, since there are $\phi'' \in \clAlg{}$ that satisfy these conditions and the construction of $\phi''$ is always from shorter to longer sequences.
    
{Define $y_i = \phi_i(\pi_\textup{longshort}) - \phi_i(\pi_\textup{longshort} - \ell(\pi_\textup{longshort}))$ for all $i \in P(\pi_\textup{longshort})$. For every agent $i$ already present in the prefix, $y_i \geqslant 0$ because $\phi$ satisfies OIR. For the newly arrived agent $\ell(\pi_\textup{longshort})$, the prefix does not contain it, so $y_{\ell(\pi_\textup{longshort})} = \phi_{\ell(\pi_\textup{longshort})}(\pi_\textup{longshort}) \geqslant 0$ because $\phi$ is nonnegative; this is exactly where the nonnegativity assumption of \Cref{thm:properties_iff_seqsh} is used, and without it a rule could pay a negative share to an arriving agent and fall outside \clAlg{}. Here} $\ell(\pi)$ is the last agent of the sequence $\pi$ and the subtraction $\pi - \ell(\pi)$ denotes that the last agent of $\pi$ is removed from the end of that sequence. Since $\phi$ satisfies I4OA, $\phi_i(\pi_\textup{longshort}) \leqslant \phi_i(\pi^*)$ for all $i \in P(\pi_\textup{longshort})$.
    Now, set $\forall i\in P(\pi_\textup{longshort})$ 
    \begin{align*}
        \Tilde{\phi}_{i}(\pi_\textup{longshort}) &= \Tilde{\phi}_{i}(\pi_\textup{longshort} - \ell(\pi_\textup{longshort})) + y_{i} \\
        &= \phi_{i}(\pi_\textup{longshort} - \ell(\pi_\textup{longshort})) + y_{i} \\
        &= \phi_{i}(\pi_\textup{longshort}).
    \end{align*}
    Note that this retains the sequential solution concept $\Tilde{\phi}$ in \clAlg{} since it adds $y_i$ on top of the previously allocated reward to the sequence $\pi_\textup{longshort} - \ell(\pi_\textup{longshort})$, since the value that $y_i$ can take is according to the function \impr. After adding $y_{\ell(\pi_\textup{longshort})}$ it gives $\phi_{\ell(\pi_\textup{longshort})}(\pi_\textup{longshort})$ which is at most $\phi_{\ell(\pi_\textup{longshort})}(\pi^*) = \Tilde{\phi}_{\ell(\pi_\textup{longshort})}(\pi^*)$. The second equality above comes because by construction $\Tilde{\phi}$ matches $\phi$ for all sequences of length less than that of $\pi_\textup{longshort}$. 
    
    Since $\pi_\textup{longshort}$ was chosen arbitrarily among all possible longest shortest sequences, and this process of assigning $\Tilde{\phi}_{i}(\pi_\textup{longshort})$ does not affect other sequences of the same length, we can do the same for all other sequences of the same length simultaneously. Hence, $\Tilde{\phi}$ matches $\phi$ for all sequences of that length and lower. This completes the proof.
\end{proof}

{
\subsection{Proof of \Cref{prop:runtime}}
\label{app:runtime}
\begin{proof}
We bound the cost of each part of \Cref{algo:SeqSh} separately and then add
the bounds.  Throughout, we assume that the worth table
$\{v(\pi)\}_{\pi \in \Pi}$ is stored so that $v(\pi)$ can be read in constant
time, and that $\phi(\pi)$ is stored as a vector of length $n$.

\emph{Step 1: the size of $\Pi$.}  A sequence of length $k$ is an ordered
choice of $k$ distinct agents out of $n$, and there are $n!/(n-k)!$ such
choices.  Summing over $k = 1, \ldots, n$ and substituting $m = n-k$ gives
\[
  |\Pi| \;=\; \sum_{k=1}^{n} \frac{n!}{(n-k)!}
       \;=\; \sum_{m=0}^{n-1} \frac{n!}{m!}
       \;=\; \Bigl(\sum_{m=0}^{n} \frac{n!}{m!}\Bigr) - 1 .
\]
By the classical identity $\sum_{m=0}^{n} n!/m! = \lfloor e \cdot n! \rfloor$
for $n \geqslant 1$, which follows from the Taylor series of $e$ because the
tail $\sum_{m > n} n!/m!$ lies strictly between $0$ and $1$, this proves
\Cref{eq:Pi-size}.  In particular $|\Pi| \geqslant n!$, which is why every
bound below that is polynomial in $|\Pi|$ is exponential in $n$.

\emph{Step 2: finding $\pi^*(v)$.}  Line~1 of \Cref{algo:SeqSh} computes a
maximizer of $v$ over $\Pi$.  Because $v$ is monotone (\Cref{sec:prelim}),
some full sequence attains the maximum, so it suffices to scan the $n!$ full
sequences, and in any case a single scan of the whole table costs
$O(|\Pi|)$ time.

\emph{Step 3: the \basis{} subroutine.}  The basis conditions
(\Cref{eq:basis}) consist of one inequality for every $\pi \in \Pi$ and one
equality for $\pi^*(v)$, all in the $n$ unknowns $x_1, \ldots, x_n$.
Deciding feasibility and returning a feasible $x$ is therefore a linear
feasibility problem with $n$ variables and $|\Pi| + 1$ constraints, whose
cost we denote by $T_{\mathrm{LP}}(n, |\Pi|)$.  By the polynomial-time
solvability of linear programming (for example, by interior-point methods),
$T_{\mathrm{LP}}(n, |\Pi|)$ is polynomial in $n$ and $|\Pi|$, and hence
polynomial in $|\Pi|$ since $n \leqslant |\Pi|$.

\emph{Step 4: initialization.}  Setting $\phi(\pi) = \mathbf{0}$ for every
$\pi \neq \pi^*(v)$, copying $x$ into $\phi(\pi^*(v))$, and setting
$\phi_i(i) = v(i)$ for every $i \in N$ touches each stored vector once and
costs $O(n\,|\Pi|)$ time.

\emph{Step 5: the arrival loop.}  The loop visits every pair $(\pi, j)$ with
$|\pi| \leqslant n-1$ and $j \in N \setminus P(\pi)$, and each such pair
corresponds to exactly one sequence $\pi + j$ of length at least two.
Conversely, every sequence of length at least two arises from exactly one such
pair, namely by removing its last agent.  Hence the loop body runs exactly
$|\Pi| - n$ times.

It remains to bound the cost of one call to \impr{}.  We implement it by a
greedy rule.  Let $M = v(\pi+j) - v(\pi)$ be the marginal worth, and let
$s_i = \phi_i(\pi^*) - \phi_i(\pi)$ be the slack of agent $i \in P(\pi+j)$.
Visiting the agents of $P(\pi+j)$ in any fixed order, the rule sets
$y_i = \min\{s_i,\; M - \sum_{i' \text{ visited before } i} y_{i'}\}$.  Each
$y_i$ is computed in constant time from a running sum, so one call costs
$O(n)$ time, and updating $\phi(\pi+j)$ from $\phi(\pi)$ and $y$ costs another
$O(n)$.  The greedy rule returns a valid output whenever one exists, because
it stops only after the running sum reaches $M$ or after all slack is used.
The feasibility argument in the proof of \Cref{thm:properties_iff_seqsh}
shows that the total slack $\sum_i s_i$ is at least $M$, by condition~(i) of
the basis conditions together with SE for $\pi$, so the running sum always
reaches $M$.  Each $s_i$ is nonnegative by I4OA for $\pi$, and $M$ is
nonnegative by monotonicity of $v$, so every $y_i$ lies in $[0, s_i]$ as
\impr{} requires.  The total cost of the loop is therefore
$O(n\,(|\Pi| - n)) = O(n\,|\Pi|)$.

\emph{Combining the steps.}  Adding the costs of Steps 2 to 5 gives a total
of $T_{\mathrm{LP}}(n, |\Pi|) + O(|\Pi|) + O(n\,|\Pi|) + O(n\,|\Pi|)$, which is
$T_{\mathrm{LP}}(n, |\Pi|) + O(n\,|\Pi|)$, as claimed.  Finally, for the
online statement, note that once $x$ has been computed in the offline phase,
the reward vector after each new arrival depends only on the current
$\phi(\pi)$, on $x$, and on the two worths $v(\pi)$ and $v(\pi+j)$.  It is
therefore produced by a single greedy call in $O(n)$ time, without
enumerating any other sequence.
\end{proof}
}

\subsection{A Worked Example of \textsc{SeqShare}}
\label{app:walkthrough}

\Cref{fig:buckets} illustrates the reward-assignment logic of
\clAlg{} on a single four-player run. To complement it, we give here a
complete numerical trace of \Cref{algo:SeqSh} on a three-agent TCG,
showing the vector returned by \textsc{Basis} and the vector $y$ returned by
\textsc{Improvize} at every arrival step.

\paragraph{The game.}
Let $N = \{1,2,3\}$ with the characteristic function in Table~\ref{tab:wt-game}.

\begin{table}[h]
\centering
\caption{A three-agent TCG. Worth of every sequence $\pi \in \Pi$.}
\label{tab:wt-game}
\begin{tabular}{lccc|cccccc|cccccc}
\toprule
$\pi$ & $1$ & $2$ & $3$ & $12$ & $13$ & $21$ & $23$ & $31$ & $32$
      & $123$ & $132$ & $213$ & $231$ & $312$ & $321$ \\
$v(\pi)$ & 2 & 1 & 1 & 4 & 3 & 3 & 2 & 3 & 2 & 6 & 5 & 5 & 4 & 5 & 4 \\
\bottomrule
\end{tabular}
\end{table}

The game is monotone: $v(\pi') \le v(\pi)$ for every prefix $\pi' \sqsubset \pi$.
Order matters throughout; for instance $v(12) = 4$ while $v(21) = 3$, and
$v(123) = 6$ while $v(321) = 4$.

\paragraph{Line 1: computing $\pi^*$ and the basis.}
The optimal sequence is $\pi^*(v) = 123$ with $v(\pi^*) = 6$. The basis
conditions of Equation~\eqref{eq:basis} require $x \in \mathbb{R}^3$ with
$x_1 + x_2 + x_3 = 6$ and $\sum_{i \in P(\pi)} x_i \ge v(\pi)$ for all
$\pi \in \Pi$. Collecting the binding constraints:
\[
x_1 \ge 2, \quad x_2 \ge 1, \quad x_3 \ge 1, \qquad
x_1 + x_2 \ge 4, \quad x_1 + x_3 \ge 3, \quad x_2 + x_3 \ge 2 .
\]
Note that each pairwise constraint is generated by the \emph{larger} of the two
orderings of that pair, e.g.\ $x_1 + x_2 \ge \max\{v(12), v(21)\} = 4$.

The feasible set is non-empty, so the game is basis-satisfiable. It is also not a
singleton: $x = (3,2,1)$, $x = (2,2,2)$, and $x = (4,1,1)$ are all valid basis
solutions. This is the first source of non-uniqueness of \textsc{SeqShare}
noted in \Cref{sec:seqsh}. We fix $x = (3,2,1)$ for the remainder of
the trace, so $\phi(\pi^*) = \phi(123) = (3,2,1)$.

\paragraph{Line 5: length-one sequences.}
Each agent arriving alone receives the full worth it generates,
$\phi_i(i) = v(i)$:
\[
\phi(1) = (2,0,0), \qquad \phi(2) = (0,1,0), \qquad \phi(3) = (0,0,1).
\]
Each is already within its cap, since $2 \le x_1$, $1 \le x_2$, $1 \le x_3$.
Observe that agent $3$ is \emph{exactly} at its cap, $\phi_3(3) = 1 = x_3$. Its
remaining slack is zero, so agent $3$ receives nothing further in any sequence
beginning with $3$. This is precisely the situation depicted in
\Cref{fig:buckets}(d).

\paragraph{Lines 7--15: the arrival loop.}
At each step, \textsc{Improvize} returns a $y$ satisfying
$y_i \in [0,\; \phi_i(\pi^*) - \phi_i(\pi)]$ and
$\sum_{i \in P(\pi + j)} y_i = v(\pi + j) - v(\pi)$, and the new allocation is
$\phi_i(\pi + j) = \phi_i(\pi) + y_i$. Table~\ref{tab:wt-trace} lists every call.

\begin{table}[h]
\centering
\caption{Trace of \Cref{algo:SeqSh} with $x = (3,2,1)$. The
``slack'' column is the componentwise upper bound $x - \phi(\pi)$ available to
\textsc{Improvize}; the marginal is $v(\pi + j) - v(\pi)$.}
\label{tab:wt-trace}
\begin{tabular}{cccccc}
\toprule
$\pi$ & $j$ & $\phi(\pi)$ & slack $x - \phi(\pi)$ & marginal & $y$ \\
\midrule
\multicolumn{6}{l}{\emph{$k = 2$}} \\
$1$ & $2$ & $(2,0,0)$ & $(1,2,1)$ & $2$ & $(1,1,0)$ \\
$1$ & $3$ & $(2,0,0)$ & $(1,2,1)$ & $1$ & $(0,0,1)$ \\
$2$ & $1$ & $(0,1,0)$ & $(3,1,1)$ & $2$ & $(2,0,0)$ \\
$2$ & $3$ & $(0,1,0)$ & $(3,1,1)$ & $1$ & $(0,1,0)$ \\
$3$ & $1$ & $(0,0,1)$ & $(3,2,0)$ & $2$ & $(2,0,0)$ \\
$3$ & $2$ & $(0,0,1)$ & $(3,2,0)$ & $1$ & $(0,1,0)$ \\
\midrule
\multicolumn{6}{l}{\emph{$k = 3$}} \\
$12$ & $3$ & $(3,1,0)$ & $(0,1,1)$ & $2$ & $(0,1,1)$ \\
$13$ & $2$ & $(2,0,1)$ & $(1,2,0)$ & $2$ & $(1,1,0)$ \\
$21$ & $3$ & $(2,1,0)$ & $(1,1,1)$ & $2$ & $(1,0,1)$ \\
$23$ & $1$ & $(0,2,0)$ & $(3,0,1)$ & $2$ & $(2,0,0)$ \\
$31$ & $2$ & $(2,0,1)$ & $(1,2,0)$ & $2$ & $(1,1,0)$ \\
$32$ & $1$ & $(0,1,1)$ & $(3,1,0)$ & $2$ & $(2,0,0)$ \\
\bottomrule
\end{tabular}
\end{table}

Three steps in Table~\ref{tab:wt-trace} are worth highlighting.

\emph{(i) Zero slack forces exclusion.} At $\pi = 13$, $j = 2$, agent $3$ has
slack $x_3 - \phi_3(13) = 0$, so $y_3 = 0$ is forced and the marginal $2$ is
split between agents $1$ and $2$ only. The same occurs at $\pi = 31$ and
$\pi = 32$. This is the mechanism behind \Cref{fig:buckets}(d).

\emph{(ii) The final step into $\pi^*$ is uniquely determined.} At $\pi = 12$,
$j = 3$, the slack is $(0,1,1)$, whose components sum to
$v(\pi^*) - v(12) = 2$, which is exactly the required marginal. Hence
$y = (0,1,1)$ is the only admissible choice. This holds in general: sequential
efficiency forces $\sum_i \phi_i(\pi) = v(\pi)$, so the total slack on entering
$\pi^*$ always equals the final marginal, and \textsc{Improvize} necessarily
recovers $\phi(\pi^*) = x$. The initialization on line 3 and the loop are
therefore consistent, never conflicting.

\emph{(iii) Freedom elsewhere.} At $\pi = 1$, $j = 2$ the slack is $(1,2,1)$ and
the marginal is $2$, so $y = (0,2,0)$ is equally admissible and yields
$\phi(12) = (2,2,0)$ instead of $(3,1,0)$. This is the second source of
non-uniqueness: \textsc{Improvize} returns a set, not a point.

\paragraph{The resulting sequential solution concept.}
Table~\ref{tab:wt-phi} gives the full $\phi$ produced by this run.

\begin{table}[h]
\centering
\caption{The sequential solution concept $\phi \in \textsc{SeqShare}$ obtained above.}
\label{tab:wt-phi}
\begin{tabular}{lcccccccc}
\toprule
$\pi$ & $1$ & $2$ & $3$ & $12$ & $13$ & $21$ & $23$ & $31$ \\
$\phi(\pi)$ & $(2,0,0)$ & $(0,1,0)$ & $(0,0,1)$ & $(3,1,0)$ & $(2,0,1)$
            & $(2,1,0)$ & $(0,2,0)$ & $(2,0,1)$ \\
\midrule
$\pi$ & $32$ & $123$ & $132$ & $213$ & $231$ & $312$ & $321$ & \\
$\phi(\pi)$ & $(0,1,1)$ & $(3,2,1)$ & $(3,1,1)$ & $(3,1,1)$ & $(2,2,0)$
            & $(3,1,1)$ & $(2,1,1)$ & \\
\bottomrule
\end{tabular}
\end{table}

All three axioms can be read off directly. \textbf{SE:} the entries of
$\phi(\pi)$ sum to $v(\pi)$ for each of the fifteen sequences.
\textbf{I4OA:} every coordinate is bounded by $x = (3,2,1)$; agent $1$ never
exceeds $3$, agent $2$ never exceeds $2$, agent $3$ never exceeds $1$.
\textbf{OIR:} along every prefix chain rewards are non-decreasing, e.g.
{$\phi_1(1) = 2 \le \phi_1(12) = 3 \le \phi_1(123) = 3$ and $\phi_2(2) = 1 \le \phi_2(23) = 2 \le \phi_2(231) = 2$}.

\section{Appendix to \Cref{sec:Shapley-solutions}}

\subsection{Proof of \Cref{lemma:reduced-property-satisfaction}}
\begin{proof} Part (a): since $\phi$ satisfies SE, we have $\sum_{i \in P(\pi)}\phi_i(\pi,v) = v(\pi), \forall \pi \in \Pi$.
Averaging this over all full-length sequences, we get $\forall \pi \in \Pi$
\begin{align*}
    \frac{1}{n!}\sum_{\pi \in \Pi: P(\pi)=N}v(\pi) &= \frac{1}{n!}\sum_{\pi \in \Pi: P(\pi)=N} \sum_{i \in P(\pi)}\phi_i(\pi,v) \\
    &=\sum_{i \in N} \left(\frac{1}{n!}\sum_{\pi \in \Pi: P(\pi)=N} \phi_i(\pi,v) \right) \\
    &= \sum_{i \in N} \overline{\phi}_i(\pi,v)
\end{align*}
The first equality comes via changing the order of the summation, and the second equality is from the definition of the reduced solution concept. Hence, $\overline{\phi}$ satisfies EE.

\medskip \noindent 
Part (b): since $\phi$ satisfies SA, we have $\phi_i(\pi,u) + \phi_i(\pi,w) = \phi_i(\pi,u+w) , \forall i \in N$. 
Averaging this over all full length sequences, we get
\begin{align*}
    & \frac{1}{n!}\sum_{\pi \in \Pi: P(\pi)=N}(\phi_i(\pi,u) + \phi_i(\pi,w)) \\&= \frac{1}{n!}\sum_{\pi \in \Pi: P(\pi)=N}\phi_i(\pi,u+w) , \: \forall \pi \in \Pi, i \in N \\
    &\implies \overline{\phi}_i(u) + \overline{\phi}_i(w) = \overline{\phi}_i(u+w) , \forall i \in N
\end{align*}
The implication holds from the definition of the reduced solution concept. Hence, $\overline{\phi}$ satisfies EA.

\medskip \noindent 
Part (c): since $\phi$ satisfies SNP, we have $v(\pi+i) = v(\pi) , \: \forall \pi \in \Pi:i \notin P(\pi) \implies \phi_i(\pi,v) = 0, \forall \pi \in \Pi$. For every such player $i$, averaging over all possible full length sequences, we get $\frac{1}{n!}\sum_{\pi \in \Pi: P(\pi)=N}\phi_i(\pi,v) = 0$, which is $\overline{\phi}_i(v) = 0$ by definition. Hence, 
\[v(\pi+i) = v(\pi) , \: \forall \pi \in \Pi:i \notin P(\pi) \implies \overline{\phi}_i(v) = 0.\]
Hence, $\overline{\phi}$ satisfies ENP.
\end{proof}

\subsection{Proof of \Cref{lemma:carrier_games_basis}}
\begin{proof}
    Consider an arbitrary TCG \seqgame{}. We show that every $v$ can be written as a linear combination of $u_{\pi}$ (\Cref{def:carrier}).

    \medskip \noindent
    {\em Part 1:} we show that $u_\pi, \pi \in \Pi$ are linearly independent. Suppose not, then $\exists \{\alpha_{\pi}:\pi\in \Pi\}$, not all $0$ such that $\sum_{\pi \in \Pi} \alpha_{\pi}u_{\pi}(\pi') = 0, \forall \pi' \in \Pi$.
    
    Let $T = \{\alpha_\pi: \pi \in \Pi, \alpha_\pi \neq 0\}$ be the set of all non-zero $\alpha_\pi$'s. Let $\tau= \{\pi:\pi \in \Pi,\alpha_\pi\in T\}$ be the indices of $\alpha$ in $T$. Observe that, any $\pi \notin \tau$ must have $\alpha_\pi=0$. Consider the permutation $\pi_0 \in \tau$, such that no prefix of $\pi_0$ is in $\tau$. We call this a {\em minimal sequence} of $\tau$. We can write  
    \begin{align*}
    \sum_{\pi \in \Pi} \alpha_{\pi}u_{\pi}(\pi_0) =& \sum_{\pi \in \Pi,\pi \sqsubset \pi_0} \alpha_{\pi}u_{\pi}(\pi_0) + 
    \alpha_{\pi_0}u_{\pi_0}(\pi_0)\\&+ \sum_{\pi \in \Pi, \pi \not\sqsubset \pi_0} \alpha_{\pi}u_{\pi}(\pi_0) = 0.
    \end{align*}

    Note that the first sum in the expanded form does not have any $\pi \in \tau$; so all $\alpha_\pi = 0$ for that sum. In the third sum, $\pi$ is not a prefix of $\pi_0$. Hence by definition of $u_\pi$, $u_{\pi}(\pi_0)=0$ for them. So we conclude that $\alpha_{\pi_0}u_{\pi_0}(\pi_0) = 0$.
    But $u_{\pi_0}(\pi_0) = 1$ by definition, and $\alpha_{\pi_0} \neq 0$ since $\pi_0 \in \tau$
    which is a contradiction.
    Hence, $u_\pi$'s are linearly independent.

    \medskip \noindent
    {\em Part 2:} we next prove that $u_\pi$ span the entire space of $\mathbb{R}^{|\Pi|}$.
    Every sequence $\pi \in \Pi$ denotes a unique carrier game $\langle N,u_{\pi} \rangle$. Hence, the number of unique $u_{\pi}$ {is} $|\Pi|$. %
 
    $u_{\pi}: \Pi \to \mathbb{R}$, so if they are linearly independent, they must span the entire vector space of $|\Pi|$ dimensions, i.e., $\mathbb{R}^{|\Pi|}$.

    Note that from parts 1 and 2, the lemma is immediate, since for any arbitrary TCG \seqgame{}, the worth function $v$ lives in $\mathbb{R}^{|\Pi|}$, since $v: \Pi \to \mathbb{R}$. For a fixed $N$, $\Pi$ is finite. Parts 1 and 2 showed that $u_\pi, \pi \in \Pi$ forms a basis of $\mathbb{R}^{|\Pi|}$, and hence any $v$ can be written as a linear combination of $u_\pi$. Therefore, every TCG \seqgame{} is a linear combination of the carrier games.
\end{proof}

\subsection{Proof of \Cref{lemma:margsol_carrier}}
\begin{proof}
 Suppose $i \in N \setminus \{\ell(\pi)\}$ is not the last player of $\pi$. In $\langle N,u_{\pi,\alpha} \rangle$, every such player satisfies $u_{\pi,\alpha}(\pi'+i) = u_{\pi,\alpha}(\pi'), \ \forall \pi' \in \Pi_{-i}$, i.e., each such $i$ is a sequential null player. Until player $\ell(\pi)$ arrives in the sequence $\pi$, no marginal contribution is generated in this TCG by any other player arriving after. Hence, by SNP, all other players must get $\phi_i(\pi',u_{\pi,\alpha}) = 0 , \forall \pi' \in \Pi$. The entire worth of $\alpha$ then must go to $\ell(\pi)$ due to SE.
\end{proof}

\subsection{Proof of \Cref{Thm:MargSol_unique}}
\begin{proof}
($\Leftarrow$) It is trivial to see that if $i$ is a sequential null player, $\MargSol_i(\pi)=0, \forall \pi \in \Pi$. Also, $\MargSol_i(\pi,u)+\MargSol_i(\pi,v) = u(\pi_i+i)+v(\pi_i+i)-u(\pi_i)-{v(\pi_i)} = \MargSol_i(\pi,u+v)$. %
Hence \MargSol{} satisfies SA and SNP. To see that it also satisfies SE, note that 
\begin{align*}
    \lefteqn{\sum_{i \in P(\pi)} \MargSol_i(\pi) = \sum_{i \in P(\pi)} \left(v(\pi_i+i)-v(\pi_i)\right)} \\
    &=(v(\pi(1))-v(\emptyset)) + (v(\pi(1)\pi(2))-v(\pi(1))) \\&+ (v(\pi(1)\pi(2)\pi(3))-v(\pi(1)\pi(2))) +  \\
    & \ldots + (v(\pi(1)\pi(2)\pi(3)\cdots\pi(|P(\pi)|))\\&-v(\pi(1)\pi(2)\cdots\pi(|P(\pi)|-1))) = v(\pi).
\end{align*}
{Here, $\pi(k)$ is the agent at the $k^\textup{th}$ position of $\pi$, and $v(\emptyset)=0$ by the convention of \Cref{sec:prelim}.}

\smallskip \noindent
($\Rightarrow$)
    Consider an arbitrary TCG \seqgame{}. Write it as a sum of its carrier games which is possible due to \Cref{lemma:carrier_games_basis}:    
\begin{equation}
    \label{eq:basis-sum}
    v(\pi') = \sum_{\pi \in \Pi} \alpha_{\pi}u_{\pi}(\pi') = \sum_{\pi \in \Pi} u_{\pi,\alpha_{\pi}}(\pi').
\end{equation}
For each carrier game, we know the structure of the unique sequential solution concept that satisfies SE and SNP from \Cref{lemma:margsol_carrier}. Note that \MargSol{} satisfies all three properties for every game \seqgame{}, hence it must satisfy them for the carrier games as well. Suppose there exists a different sequential solution concept $\phi$ that also satisfies the same three properties. We will prove that $\phi$ must be \MargSol{}.

From \Cref{lemma:margsol_carrier} and the discussion above, we conclude that $\phi$ must be the same as \MargSol{} for the carrier game $\langle N,u_{\pi,\alpha} \rangle$. Hence we have $\forall i \in N, \forall \pi, \pi' \in \Pi$
\begin{align*}
    &\MargSol_i(\pi',u_{\pi,\alpha_{\pi}}) = \phi_i(\pi',u_{\pi,\alpha_{\pi}}), \\
    &\Rightarrow \sum_{\pi \in \Pi} \MargSol_i(\pi',u_{\pi,\alpha_{\pi}}) = \sum_{\pi \in \Pi} \phi_i(\pi',u_{\pi,\alpha_{\pi}}), \\
    &\Rightarrow \MargSol_i\left(\pi',\sum_{\pi \in \Pi} u_{\pi,\alpha_{\pi}}\right) = \phi_i\left(\pi',\sum_{\pi \in \Pi} u_{\pi,\alpha_{\pi}}\right), \\
    &\Rightarrow \MargSol_i\left(\pi', v \right) = \phi_i\left(\pi', v \right).
\end{align*}
The first implication holds by summing over all $\pi \in \Pi$. The second implication is by SA for both sequential solution concepts. The third implication is by \Cref{eq:basis-sum}. Since $v$ was arbitrary, we conclude that $\phi$ is the same as \MargSol{}.
\end{proof}

\subsection{Uniqueness of \ExtSh{}}
\label{sec:unique-extsh}

\ExtSh{} is an averaged version of \MargSol{} (\Cref{eq:ExtShapley}).
Although \MargSol{} uniquely satisfies SE, SA, and SNP
(\Cref{Thm:MargSol_unique}), this does not automatically imply that \ExtSh{}
uniquely satisfies EE, EA, and ENP.  However,
\citet[Theorem~1]{NOWAK1994150} established that \ExtSh{} indeed uniquely
satisfies EE, EA, and ENP.  We restate this theorem and present an
alternative proof using our framework. 

\begin{theorem}[\citet{NOWAK1994150}]
\label{Thm:ExtSh_unique}
For every TCG \seqgame{}, an extended solution concept $\psi \in \Psi$
satisfies EE, EA, and ENP if and only if $\psi = \ExtSh{}$.
\end{theorem}
\begin{lemma}
\label{lemma:carrier_ExtSh}
Consider $\langle N, u_{\pi,\alpha} \rangle$ as in \Cref{lemma:margsol_carrier}.
If $\psi \in \Psi$ satisfies EE and ENP, then $\psi_i(u_{\pi,\alpha}) =
\frac{(n-|P(\pi)|)!}{n!}\alpha$ if $i = \ell(\pi)$ and $0$ otherwise.
\end{lemma}
\begin{proof}
    We know from \Cref{lemma:margsol_carrier} that every $i \in N \setminus \{\ell(\pi)\}$ is a null player and should get zero reward share by any extended solution concept $\psi$ by ENP. Hence, the only positive reward should go to $\ell(\pi)$. By EE, we get
    $$
    \sum_{i\in P(\pi')} \psi_i(u_{\pi,\alpha}) = \frac{1}{n!}\sum_{\pi' \in \Pi: P(\pi') = N} u_{\pi,\alpha}(\pi').
    $$
    Given the earlier arguments, this reward will go entirely to $\ell(\pi)$, i.e., 
    {for $i = \ell(\pi)$, the last player of $\pi$,} we get that $\psi_{\ell(\pi)}(u_{\pi,\alpha}) = \frac{1}{n!}\sum_{\pi' \in \Pi: P(\pi') = N} u_{\pi,\alpha}(\pi')$. 

    By definition of $u_{\pi,\alpha}$, as long as $\pi$ is a prefix of a given sequence of players $\pi'$, the remaining $N \setminus P(\pi)$ players can appear in any sequence. So,
    there are $(n-|P(\pi)|)!$ different $\pi' \in \Pi$ such that $\pi$ is a prefix of $\pi'$. Therefore, $u_{\pi,\alpha}(\pi') = \alpha$ for all those $\pi'$.  Hence,  $\psi_{\ell(\pi)}(u_{\pi,\alpha}) = \frac{(n-|P(\pi)|)!}{n!}\alpha$.
\end{proof}

\begin{proof}[Proof of \Cref{Thm:ExtSh_unique}]
($\Leftarrow$) We have already proved in \Cref{Thm:MargSol_unique} that \MargSol{} satisfies SE, SA, and SNP, and by \Cref{lemma:reduced-property-satisfaction}, its reduced solution concept \ExtSh{} satisfies EE, EA, and ENP.

\smallskip \noindent
($\Rightarrow$)
    This direction of the proof follows in similar lines to that of \Cref{Thm:MargSol_unique}. Assume that there exists an extended solution concept $\psi \neq \ExtSh{}$ that also satisfies the three given properties.

    For each carrier game, we know the structure of the unique extended solution concept that satisfies {EE and ENP} from \Cref{lemma:carrier_ExtSh}.
Note that \ExtSh{} satisfies all three properties for every game \seqgame{}, hence it must satisfy them for the carrier games as well. 
    Hence $\psi$ must be same as \ExtSh{} for the carrier game $\langle N,u_{\pi,\alpha} \rangle$. Hence we have
\begin{align*}
    &&\ExtSh_i(u_{\pi,\alpha_{\pi}}) &= \psi_i(u_{\pi,\alpha_{\pi}}), \ \forall i \in N, \forall \pi \in \Pi, \\
    &\Rightarrow& \sum_{\pi \in \Pi} \ExtSh_i(u_{\pi,\alpha_{\pi}}) &= \sum_{\pi \in \Pi} \psi_i(u_{\pi,\alpha_{\pi}}), \ \forall i \in N, \\
    &\Rightarrow& \ExtSh_i\left(\sum_{\pi \in \Pi} u_{\pi,\alpha_{\pi}}\right) &= \psi_i\left(\sum_{\pi \in \Pi} u_{\pi,\alpha_{\pi}}\right), \ \forall i \in N, \\
    &\Rightarrow& \ExtSh_i\left(v \right) &= \psi_i\left(v \right), \ \forall i \in N.
\end{align*}
The first implication holds by summing over all $\pi \in \Pi$. The second implication is by EA for both extended solution concepts. The third implication is by \Cref{eq:basis-sum}. Since $v$ was arbitrary, we conclude that $\psi$ is the same as \ExtSh{}.
\end{proof}

\section{Why Symmetry Is Not Necessary}
\label{sec:symmetry-not-necessary}

Properties in the TCG framework are more restrictive, as they make claims for
all \emph{sequences} rather than all \emph{sets} as in classical cooperative
games.  Hence, three properties suffice to ensure unique solution concepts.
In the following, we show that \MargSol{} and \ExtSh{} satisfy SS
(\Cref{def:seq_symmetry}) and ES (\Cref{def:ext_symmetry}) respectively.

\begin{theorem}
\label{thm:symmetry-satisfied}
\MargSol{} and \ExtSh{} satisfy SS and ES, respectively.
\end{theorem}
\begin{proof}
{\em Part 1 (\MargSol{})}: Consider a TCG \seqgame{}, s.t. $v(\pi)=v(\pi_{\swap{i}{j}}), \ \forall \pi \in \Pi$. We show the following $\forall \pi \in \Pi, \forall i \in N$
\begin{align*}
    \lefteqn{\MargSol_i(\pi)} \\
    &= v(\pi_i+i)-v(\pi_i) \\
    &= v((\pi_i)_{\swap{i}{j}}+j) - v((\pi_i)_{\swap{i}{j}}) \\
    &= v((\pi_{\swap{i}{j}})_j+j) - v((\pi_{\swap{i}{j}})_j) \\
    &=\MargSol_j(\pi_{\swap{i}{j}}).
\end{align*}
The first equality follows from definition. The second equality comes from the `if' condition of the SS property. The third equality follows from the fact that $(\pi_i)_{\swap{i}{j}} = (\pi_{\swap{i}{j}})_j$. The last equality is again by definition. Hence, we show that \MargSol{} satisfies SS.

\smallskip \noindent
{\em Part 2 (\ExtSh{})}: Since $\ExtSh{}_i = \frac{1}{n!}\sum_{\pi \in \Pi: P(\pi) = N} \MargSol_i(\pi)$, and we have already shown in Part 1 that $\MargSol_i(\pi) = \MargSol_j(\pi_{\swap{i}{j}})$, it is straightforward to show that 
\begin{align*}
    \lefteqn{\ExtSh{}_i} \\
    &= \frac{1}{n!}\sum_{\pi \in \Pi: P(\pi) = N} \MargSol_i(\pi) \\
    &= \frac{1}{n!}\sum_{\pi \in \Pi: P(\pi) = N} \MargSol_j(\pi_{\swap{i}{j}}) \\
    &= \frac{1}{n!}\sum_{\pi \in \Pi: P(\pi) = N} \MargSol_j(\pi) \\
    &= \ExtSh{}_j.
\end{align*}
The third equality holds since $\pi_{\swap{i}{j}}$ just changes the order of summation in the set $\{\pi \in \Pi: P(\pi) = N\}$.
Hence, $\ExtSh$ satisfies ES.
\end{proof}

\noindent\textsc{Remark.}  In classical cooperative games, symmetry is
required to make the Shapley value unique because a carrier game is defined
with respect to a critical \emph{set} of players, and dividing worth among
multiple non-null players requires symmetry.  In TCGs, a carrier game is
defined with respect to a critical \emph{sequence}: only the last player is
non-null and receives the entire worth, so symmetry is not required.

\section{Appendix to \Cref{sec:bestofboth}}

\subsection{Proof of \Cref{lemma:extended_space_proj}}
\begin{proof}
    Consider a sequential solution concept $\phi \in \clAlg{}$ such that $\overline{\phi}=\ExtSh$.
    Suppose $\exists i \in N$ such that $\phi_i(\pi^*(v)) < \ddfrac{1}{n!}\sum_{\pi \in \Pi : P(\pi) = N} v(\pi_i+i)-v(\pi_i)$. Since $\phi \in \clAlg{}$, it satisfies I4OA, i.e.,
    \begin{align*}
        &\phi_i(\pi) \leqslant \phi_i(\pi^*(v)), \ \forall \pi \in \Pi \\
        \implies &\phi_i(\pi) < \ddfrac{1}{n!}\sum_{\pi \in \Pi : P(\pi) = N} v(\pi_i+i)-v(\pi_i), \ \forall \pi \in \Pi.
    \end{align*}
    Averaging over all possible $\pi \in \Pi$ such that $P(\pi) = N$, we get
    \begin{align}
        \overline{\phi}_i =& \ddfrac{1}{n!}\sum_{\pi \in \Pi : P(\pi) = N} \phi_i(\pi) \\ <& \ddfrac{1}{n!}\sum_{\pi \in \Pi : P(\pi) = N} v(\pi_i+i)-v(\pi_i) = \ExtSh_i.
    \end{align}
This contradicts the hypothesis that $\overline{\phi}=\ExtSh$.
\end{proof}

\subsection{Convex Games}
\label{app:convex}
In classical cooperative games, the Shapley value always lies in the core for
convex games.  Since the basis conditions (\Cref{eq:basis}) are analogous to
the core conditions, we check for a similar result in TCGs.

\begin{definition}[Convex Temporal Cooperative Games]
\label{def:convex-game}
A TCG \seqgame{} is \emph{convex} if $v(\pi+i)-v(\pi) \leqslant
v(\pi'+i)-v(\pi')$ for all {$\pi, \pi' \in \Pi$ and $i \in N$} such that
$i \notin P(\pi')$ and $P(\pi) \subseteq P(\pi')$.
\end{definition}

The answer is unfortunately negative.  First, the following example
illustrates a convex TCG with no basis solution, unlike convex classical
games where the Shapley value always lies in the core.

\begin{example}
Consider the convex TCG from \Cref{example:no-general-reduction} with
$a=7, b=4$.  The basis solution must satisfy $x_1 \geqslant 1, x_2 \geqslant
4, x_3 \geqslant 4$ and $x_1+x_2+x_3=8$, which is impossible.  Hence this
convex TCG has no basis solution.
\end{example}

\begin{example}
The TCG from \Cref{example:no-general-reduction} (with $a=7, b=3$) is convex
and violates \Cref{eq:necessary_extSh_SeqSh}.
\end{example}

\subsection{Simple Games}
\label{app:simple}

\begin{definition}[Simple Temporal Cooperative Games]
\label{def:simple_game}
A TCG \seqgame{} with $v:\Pi \rightarrow \{0,1\}$ is called a \emph{simple}
temporal cooperative game.
\end{definition}

Since we consider monotone games, $v(\pi)=1 \implies v(\pi')=1$ for all
$\pi \sqsubset \pi'$.  The following example shows that even for simple games,
there may be no solution in \clAlg{} reducing to \ExtSh{}.

\begin{example}
    Consider the simple TCG, $ v(1)=v(12)= v(123)=v(13) = v(132) =v(213) = v(231)= v(312) =  v(321) = 1, v(\pi)=0$ for all others, in particular $v(23)=0$.
    The unique basis solution (\Cref{eq:basis}) is $x_1=1 ,x_2=0, x_3=0$. 
    Therefore any $\phi \in \clAlg{}$ assigns $\phi_i(\pi^*)=x_i$. Routine calculations for Player 2 {give} %
 $\ExtSh{}_2=\frac{1}{6}$. However, $\phi_2(\pi^*) = 0 < \frac{1}{6} = \ExtSh{}_2$, which violates \Cref{eq:necessary_extSh_SeqSh}.
\end{example}

Even for simple TCGs, we cannot hope to have a solution in \clAlg{} that reduces to \ExtSh{}.

\end{document}